\documentclass[%
superscriptaddress,
 preprint,
showkeys,showpacs,preprintnumbers,
amsmath,amssymb,
aps,
prb,
]{revtex4-2}

\usepackage{graphicx}
\usepackage{dcolumn}
\usepackage{bm}

\usepackage{upgreek}
\usepackage{amsmath}

\begin{document}


\title{Four-channel Imaging Based on Reconfigurable Metasurfaces: Hyperchaotic Encryption under Physical Protection}

\author{Yifan Li}
\affiliation{School of Information Engineering, Nanchang University, Nanchang 330031, China}
\affiliation{Institute for Advanced Study, Nanchang University, Nanchang 330031, China}

\author{Yuhan Yang}
\affiliation{Jiluan Academy, Nanchang University, Nanchang 330031, China}

\author{Qiegen Liu}
\affiliation{School of Information Engineering, Nanchang University, Nanchang 330031, China}

\author{Shuyuan Xiao}
\email{syxiao@ncu.edu.cn}
\affiliation{School of Information Engineering, Nanchang University, Nanchang 330031, China}
\affiliation{Institute for Advanced Study, Nanchang University, Nanchang 330031, China}

\author{Tingting Liu}
\email{ttliu@ncu.edu.cn}
\affiliation{School of Information Engineering, Nanchang University, Nanchang 330031, China}
\affiliation{Institute for Advanced Study, Nanchang University, Nanchang 330031, China}

\begin{abstract}

Metasurfaces facilitate high-capacity optical information integration by simultaneously supporting near-field nanoprinting and far-field holography on a single platform. However, conventional multi-channel designs face critical security vulnerabilities for sensitive information due to insufficient encryption mechanisms. In this work, we propose a four-channel phase-change metasurface featuring algorithm-physical co-security—a dual-protection framework combining intrinsic metasurface physical security with chaotic encryption. Our polarization-multiplexed metasurface generates four optical imaging channels through meta-atom design, including two far-field holograms and two near-field patterns. To enhance system security, we apply Chen hyperchaotic encryption combined with the Logistic map and DNA encoding to convert near-field information into secure QR codes; far-field holograms are retained to demonstrate the metasurface's information capacity and for attack detection. Phase-change metasurface further provides physical-layer security by dynamically switching imaging channels via crystalline-to-amorphous state transitions, enhancing anti-counterfeiting and reliability. The proposed metasurface achieves high-fidelity imaging, robust anti-attack performance, and independent channel control. This integrated approach pioneers a secure paradigm for high-density optical information processing.

\end{abstract}

\maketitle


\section{Introduction}

In the era of global digitalization, information security has evolved into a critical concern impacting both individual privacy and national strategic interests \cite{xu2025chaotic,ning2025enhanced,ouyang2021underwater}. This challenge is particularly acute in optical information transmission and display systems \cite{feng2023diatomic}, where high-security and high-precision protection mechanisms is imperative. Metasurfaces—subwavelength artificial structures with unprecedented electromagnetic wave control precision \cite{yin2024multi,sun2012gradient,li2024intelligent}—have revolutionized information optics since the seminal work by Capasso et al \cite{yu2011light}. Through deliberate engineering of meta-atom geometry and orientation, they enable dynamic modulation of incident light's polarization \cite{frese2019nonreciprocal,deng2020malus,Xu2025}, intensity \cite{liu2021multifunctional,lu2024metasurface}, phase \cite{xie2021generalized,Li2025}, and wavelength \cite{zhang2019colorful,wang2016visible,huang2015aluminum,li2021full,faraji2018compact}, facilitating integrated functionalities including holographic imaging \cite{huang2013three,zheng2015metasurface,ye2016spin,wan2017metasurface}, nano-printing \cite{yue2018high,dai2020dual,deng2020multiplexed,goh2014three,dalloz2022anti,wang2020complete}, metalenses \cite{arbabi2015subwavelength,arbabi2016miniature,chen2017reconfigurable,chen2018broadband,wang2021design,tseng2021neural,zhang2023terahertz}, edge detection \cite{liu2024edge, Liu2025, Zhang2025, Zong2025}, and vortex beam generation \cite{shalaev2015high,mehmood2016visible,ren2019metasurface,bao2020minimalist,shi2023guided}. This property makes it possible to integrate different information channels onto a single metasurface \cite{arbabi2015dielectric,overvig2019dielectric,zhang2019multichannel,hu20193d,dai2020single,li2020three}, significantly enhancing the efficiency of multiplexing. Particularly for imaging applications, this enables metasurfaces to achieve superior resolution and increased information capacity. Nevertheless, this multi-channel approach introduces significant security challenges in contemporary information systems. Conventional optical imaging systems typically display information in plaintext form and inherently lack cryptographic protection, leaving them susceptible to sophisticated threats including exhaustive attacks and physical tampering techniques such as cropping.

To address these security challenges, various encryption schemes have been developed \cite{zheng2021metasurface,yu2024high,cao2022four,xing2025metasurface}. Among them, chaotic systems stand out as promising candidates for building high-security encryption architectures \cite{zhu2025metasurface}. Unlike metasurface-based strategies that primarily rely on predetermined physical responses and simple encryption steps, chaotic encryption employs dynamic algorithmic processes to establish cryptographic security, offering a vastly expandable and updatable key space through digitalized initial conditions and parameters. This method constructs a highly nonlinear and unpredictable mapping \cite{levy1994chaos} between plaintext and ciphertext, resulting in noise-like outputs that are inherently resistant to statistical analysis. As nonlinear dynamical systems, chaotically-based encoders exhibit extreme sensitivity to initial conditions and intrinsic randomness—properties that are highly advantageous for secure communications \cite{chen2015fast,yoon2010image,zhu2011chaos}. In image encryption applications specifically, chaotic systems generate highly complex key-dependent sequences that effectively resist exhaustive and cropping attacks \cite{zhou2024two}. However, implementing chaotic encryption in optical imaging faces two fundamental challenges, including the high complexity of key management systems, and stringent precision requirements for continuous grayscale value representation. These limitations necessitate a co-design approach integrating physical security mechanisms.

This paper proposes a high-security and high-capacity optical encryption metasurface, through synergistic integration of a Chen hyperchaotic system with phase-change materials. Leveraging polarization multiplexing, the metasurface simultaneously generates four distinct imaging channels in the visible spectrum, including two near-field nanoprinted QR codes and two far-field holographic images. To address critical security vulnerabilities in optical imaging, these channels' information is robustly encrypted via the hyperchaotic mechanism. A dedicated chaotic encryption optimization algorithm incorporating DNA encoding is designed to resolve grayscale precision limitations in QR codes, significantly relaxing the dependence on optical imaging precision without compromising cryptographic robustness. Through a recoding mechanism, this approach effectively enhances data tolerance to imaging noise and distortion. Independent high-quality decryption across all channels is demonstrated, with near-field nanoprinting achieving structural similarity index (SSIM) values up to 99\%, while far-field holographic channels reach SSIM values of 85.43\% and 72.04\%, respectively. The hybrid algorithm combining Chen hyperchaos with DNA encoding also provides superior key sensitivity and attack resistance compared to conventional methods. Moreover, the $\mathrm{Sb_2}$$\mathrm{S_3}$ based phase-change metasurface enables active, reversible switching of individual imaging channels via its crystalline-amorphous state transitions, providing dynamic physical-layer security that enhances anti-attack capabilities and introduces unclonable physical features. This scheme successfully integrates high-density optical information multiplexing with multi-layer algorithm-physical encryption on a single metasurface, establishing new pathways for secure optical information processing, dynamic displays, and high-dimensional data storage.

\section{Results and discussion}

As illustrated in Fig. 1, this paper implements a dual-layer security architecture that integrates four optical channels on a single metasurface. The two near-field channels project QR codes encrypted via the hyperchaotic system, converting plaintext into cryptographically-secured patterns, while the two far-field channels reconstruct distinct holographic images (a ``boat'' and the ``NCU'' emblem), demonstrating the substantial information capacity of the metasurface. Beyond their imaging function, the far-field channels function as sensitive optical indicators for security monitoring, operating analogously to chemical test papers by immediately revealing attacks on the channel. Any attack on the channels causes measurable changes in the far-field output, providing clear visual warning to the receiver and supporting rapid response. Specifically, under \textit{x}-polarized illumination in the amorphous state, the co-polarized far-field channel reconstructs the ``boat'' hologram, while the corresponding near-field channel displays a hyperchaotically-encrypted QR code. Under \textit{y}-polarized illumination, the cross-polarized far-field channel generates the ``NCU'' hologram alongside a different encrypted QR pattern in the near-field. The near-field QR codes act as ciphertexts containing hyperchaotically-encrypted data, with embedded keys enabling exclusive decryption and plaintext recovery. Additionally, the phase-change mechanism introduces reversible physical-layer security: upon transitioning to the crystalline state, all imaging functionalities are disabled, concealing both the holographic and encrypted content. This architecture establishes dual-layer security through combination between algorithmic encryption and physical security mechanisms.

\begin{figure*}[htbp]
	\centering
	\includegraphics[width=0.9\textwidth]{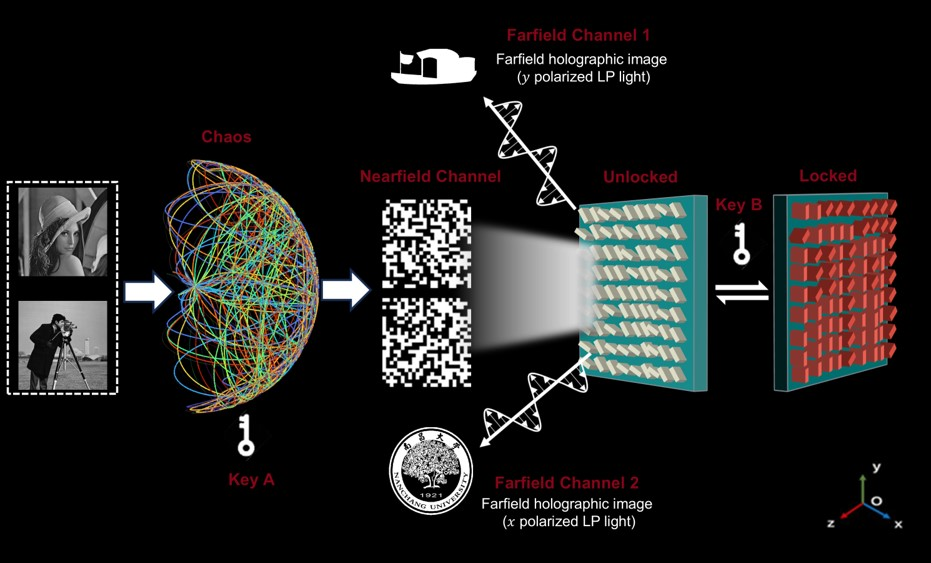}
	\caption{The metasurface generates two chaotic-encrypted QR codes in the near-field and two holograms in the far-field. Under \textit{x}-polarized light, the far-field produces a ``boat'' hologram while the near-field reveals its corresponding chaotic-encrypted QR code. Under \textit{y}-polarized illumination, the system displays an ``NCU'' emblem hologram in the far-field accompanied by another chaotic-encrypted QR code in the near-field. After transition to the crystalline state, the imaging functionality will be locked.}
	\label{fig1}	
\end{figure*}

\subsection{Metasurface design}

Typically, the optical response of a meta-atom can be described by a $2\times2$ Jones matrix as $J = \left[\begin{smallmatrix}t_{xx} & t_{xy} \\ t_{yx} & t_{yy}\end{smallmatrix}\right]$, where $x$ and $y$ represent a pair of orthogonal linear polarization states. Each matrix element $t_{ij}$ is defined as: $t_{ij} = A_{ij}e^{i\varphi_{ij}}(i,j = x,y)$, where $A_{ij}$ and $\varphi_{ij}$ represent the amplitude and phase shift of transmission coefficient from \textit{j}-polarized input light to \textit{i}-polarized output light respectively. For a single-layer metasurface, the off-diagonal components of Jones matrix are identical ($t_{xy} = t_{yx}$) due to the mirror-symmetry. It can be derived that the Jones matrix is entirely determined by five physical independent variables ($A_{x0}$, $\varphi_{x0}$, $A_{y0}$, $\varphi_{y0}$, $\theta$). Consequently, the multiplexing capability using the Jones matrix of a single anisotropic meta-atom is fundamentally limited to a maximum of five channels—three amplitude channels and two phase channels. In previous attempts \cite{li2025tunable,liu2022phase}, we found that although the single-cell structure of the metasurface can theoretically meet the requirements of four channels, but it inevitably produces unavoidable noise spots and crosstalk during imaging, and thus cannot achieve the four-channel high-resolution imaging required in this paper. To overcome the inherent limitations of single-cell metasurfaces in achieving pixel-level precision for multi-channel QR code imaging, we decide to use double-cell structure. Assume that there is no near-field coupling between adjacent nanopillars, the total output light is produced by the interference of the output light from each individual nanopillar as: $E_{\mathrm{out}} = \sum_{k = 1}^{n}E_{\mathrm{out}_{k}}$, where $k$ is the index of the $k^{\mathrm{th}}$ nanopillar. Therefore, the Jones matrix of the meta-atom composed of $k$ nanopillars with respective orientation angle $\theta_{k}$ is:

\begin{equation}
	\begin{split}
		J = \sum_{k = 1}^{n}J_{k} = \sum_{k = 1}^{n}{R^{-1}(\theta_{k})}
		\begin{bmatrix}
			A_{x_{k}}e^{i\varphi_{x_{k}}} & 0 \\
			0 & A_{y_{k}}e^{i\varphi_{y_{k}}}
		\end{bmatrix} R(\theta_{k})
	\end{split}
\end{equation}
where $J_{k}$ is the Jones matrix of the $k$\textsuperscript{th} nanopillar. Here, we fix the amplitude variables to 1 to ensure that the meta-atom presents maximum transmission, and derive the relationship between the variables and the Jones matrix elements in a double-cell meta-atom as:

\begin{equation}
	\begin{cases}
		t_{xx} = A_{xx}e^{i\varphi_{xx}} = \frac{1}{2}\sum_{k = 1}^{2}\left(e^{i\varphi_{x_{k}}}\cos^{2}\left( \theta_{k} \right) + e^{i\varphi_{y_{k}}}\sin^{2}\left( \theta_{k} \right)\right) \\
		t_{xy} = A_{xy}e^{i\varphi_{xy}} = \frac{1}{2}\sum_{k = 1}^{2}\left(e^{i\varphi_{y_{k}}} - e^{i\varphi_{x_{k}}}\right)\sin\left(\theta_{k}\right)\cos\left(\theta_{k}\right) \\
		t_{yy} = A_{yy}e^{i\varphi_{yy}} = \frac{1}{2}\sum_{k = 1}^{2}\left(e^{i\varphi_{x_{k}}}\sin^{2}\left( \theta_{k} \right) + e^{i\varphi_{y_{k}}}\cos^{2}\left( \theta_{k} \right)\right)
	\end{cases} 
\end{equation}

By selecting the $t_{xx}$ and $t_{xy}$ terms for four-channel metasurface control, we can independently modulate both the amplitude and phase of these two Jones matrix components. Based on Eq. (2), numerical solutions for the four independent variables—phases and orientation angles of the two nanopillars—can be efficiently obtained to match the desired objective functions. All modulation is achieved under zero-order diffraction, ensuring high optical efficiency and spatial resolution in the four-channel metasurface.

The metasurface's optical response is engineered through precise size and orientation design of individual meta-atom. The design process of the metasurface is as follows: Initially, a library of propagation phases ($\varphi_{x}$, $\varphi_{y}$) and transmission coefficients ($T_{x}$, $T_{y}$) is constructed by systematically varying the geometric parameters of the nanopillars. The dimensions and orientation angles of nanopillars are then optimized using the Adam algorithm in Python to achieve the desired phase profiles and high transmission performance. Finally, the two chosen nanopillars are rotated by their respective angles $\theta_{1}$ and $\theta_{2}$ and positioned diagonally within a shared square unit cell. As depicted in Fig. 2(a), we use a double-cell nanostructure comprising  $\mathrm{Sb_2}$$\mathrm{S_3}$ nanobricks on a $\mathrm{SiO_2}$ substrate. The nanobrick has a period \textit{C} = 720 nm and height \textit{H} = 1000 nm. To achieve high-fidelity imaging, the nanostructures must simultaneously possess high transmittance and full 0-2$\pi$ phase coverage at the operational wavelength. Figs. 2(c), (d) demonstrate that the nanopillars satisfy these requirements at 633 nm, providing both transmission and complete phase modulation. The phase-change material $\mathrm{Sb_2}$$\mathrm{S_3}$ has crystalline and amorphous states with significant property differences between them. In Figs. 2(e), (f), we characterize the transmission and propagation phases in the crystalline state, revealing pronounced distinctions from those in the amorphous state. We utilize this difference to achieve physical-layer optical encryption, making the optical encryption system more reliable.

\begin{figure*}[htbp]
	\centering
	\includegraphics[width=0.9\textwidth]{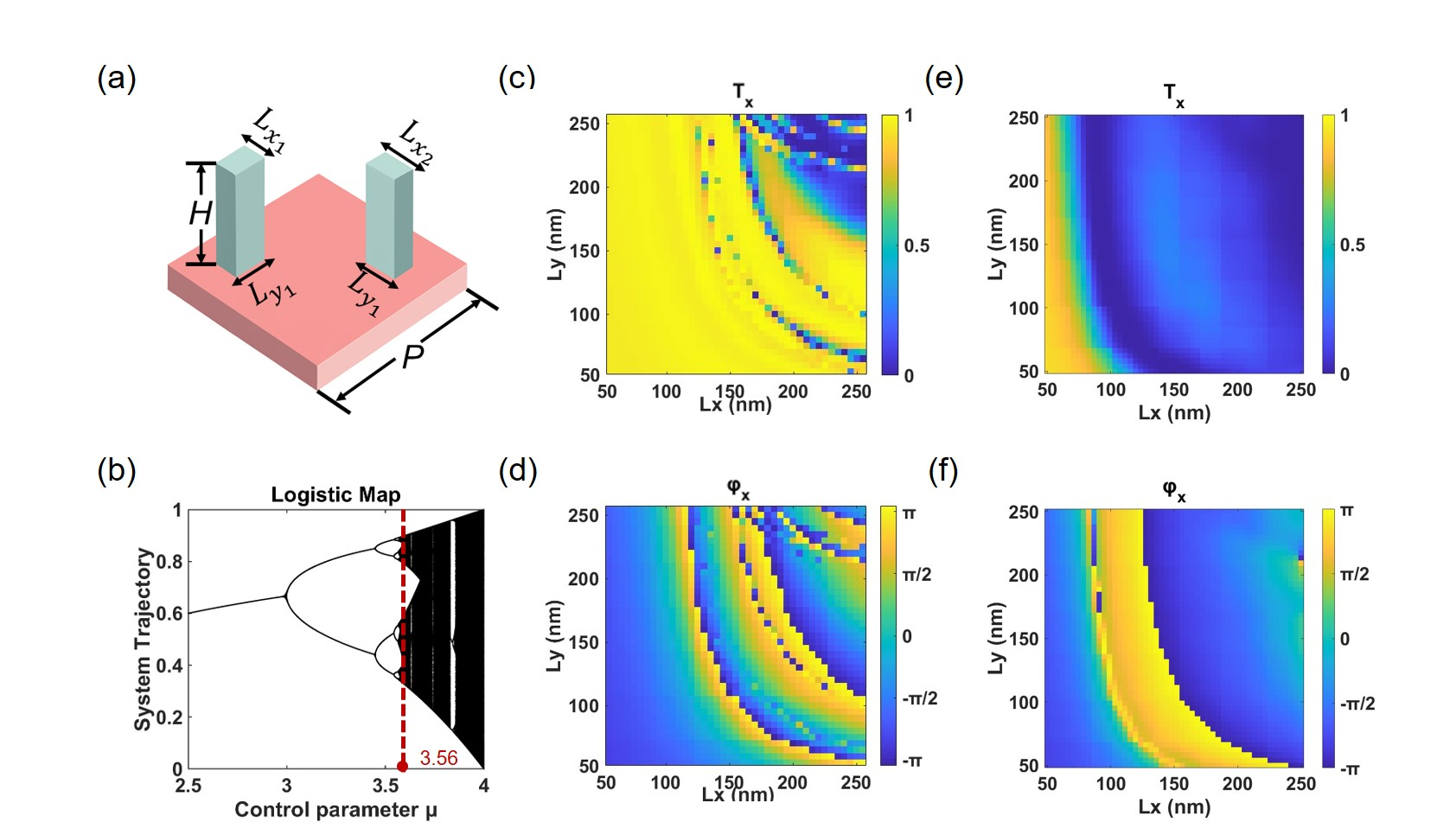}
	\caption{Metasurface nanostructure design and chaotic system overview. (a) Double-cell nanopillar structure used: $\mathrm{Sb_2}$$\mathrm{S_3}$ nanobrick on $\mathrm{SiO_2}$ substrate. (b) Diagram of the Logistic chaotic map, showing system trajectories for different control parameters $\mu$. (c), (d) Transmittance and propagation phase for long axes of double-cell nanopillars in the amorphous state. (e), (f) Propagation phase and transmittance for long axes of double-cell nanopillars in the crystalline state.}
	\label{fig2}
\end{figure*}

\subsection{Design of hyperchaotic image encryption system}

Chaotic systems, characterized by their inherent nonlinearity, exhibit seemingly random and irregular patterns despite being governed by deterministic rules. These systems possess fundamental unpredictability and exhibit high sensitivity to initial conditions, which serve as critical encryption keys. This study uses two chaotic systems for image encryption: the Logistic map and Chen hyperchaotic system. Within this framework, the initial values and control parameters function as encryption keys. Both the keys and original plaintext data are stored in QR codes. Furthermore, to augment cryptographic robustness and enhance resistance against diverse attacks, DNA encoding and DNA block operations are integrated into the proposed encryption system.
The Logistic map is a classic one-dimensional system:
\begin{equation}
	x_{n + 1} = \mu x_{n}\left( 1 - x_{n} \right) 
\end{equation}
where $x_{n}$ is the state variable, \textit{n} is the iteration number, and $\mu$ is the control parameter that determines the chaotic state of the system. As demonstrated in Fig. 2(b), chaotic behavior emerges exclusively within the parameter range  $3.56 < \mu \leq 4$. Beyond this interval, the system exhibits deterministic trajectories and lacks the cryptographic properties essential for effective encryption. In this paper, $\mu$ is set to 3.99.

Besides the Logistic map, we also introduce the Chen hyperchaotic system to our encrypted system. It is defined as:
\begin{equation}
	\begin{cases}
		\frac{dx}{dt} = a(y - x) \\
		\frac{dy}{dt} = (c - a)x - xz + cy \\
		\frac{dz}{dt} = xy - bz \\
		\frac{dw}{dt} = - dw + xz
	\end{cases} 
\end{equation}
The system consists of four state variables, denoted as $x$, $y$, $z$, and $w$, along with four control parameters $a$, $b$, $c$, and $d$. These parameters are assigned the values $a = 35$, $b = 3$, $c = 28$, and $d = 5$. In this four-dimensional configuration, the system exhibits multiple positive Lyapunov exponents, indicating the presence of hyperchaos—a regime characterized by higher dynamical complexity than conventional chaos. Such hyperchaotic behavior significantly enhances cryptographic security by providing a larger key space and greater unpredictability in system trajectories.

\begin{figure*}[htbp]
	\centering
	\includegraphics[width=0.9\textwidth]{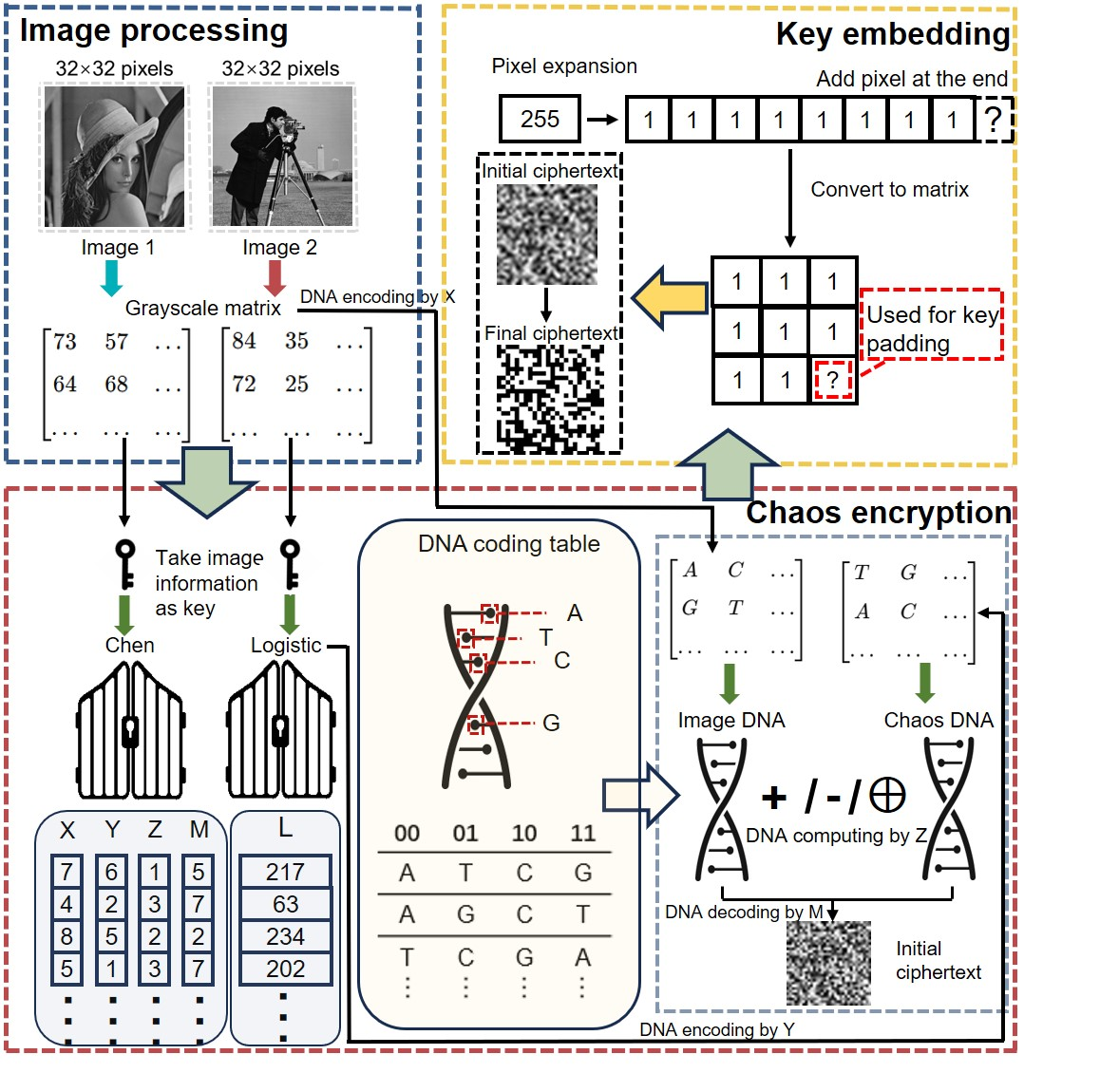}
	\caption{Chaotic encryption flowchart. The process starts by preprocessing the image and initializing the Logistic and Chen systems to generate chaotic sequences $L$, $X$, $Y$, $Z$, and $M$. The plaintext is encoded into nucleotide sequences (A, C, G, T) using DNA rules guided by $X$, then undergoes DNA operations with $L$ determined by $Z$. The results are decoded via rules from $M$ to form the initial ciphertext, which is further converted into binary and reorganized into a $3 \times 3$ matrix to produce the final QR ciphertext with embedded keys.}
	\label{fig3}
\end{figure*}

Based on the above hyperchaotic systems, we develop an image encryption algorithm. As depicted in Fig. 3, the algorithm first converts the 32×32 plaintext image into a grayscale matrix. Subsequently, it extracts key information from this matrix to initialize the chaotic system as encryption keys. This process generates five critical chaotic sequences (L, X, Y, Z, M). To substantially enhance encryption complexity and attack resistance, the algorithm innovatively introduces DNA encoding. DNA encoding dramatically expands the cryptographic key space through distinct mapping rules per nucleotide bases conversion. This inherent confusion mechanism bolsters security against statistical cryptanalysis while maintaining minimal impact on key management and decryption complexity. The algorithm converts each pixel's grayscale value into an 8-bit binary sequence. Using sequence X, it dynamically selects one of eight DNA encoding rules to map these binary sequences to nucleotide bases (A, T, C, G). Next, sequence Y controls DNA operations between the encoded image sequences and chaotic sequence L, while sequence Z determines computational operators (ADD, SUB, XOR) through value-dependent mapping. Finally, sequence M governs the decoding of processed DNA sequences back to grayscale values, generating the initial continuous ciphertext image.
\begin{figure*}[htbp]
	\centering
	\includegraphics[width=0.9\textwidth]{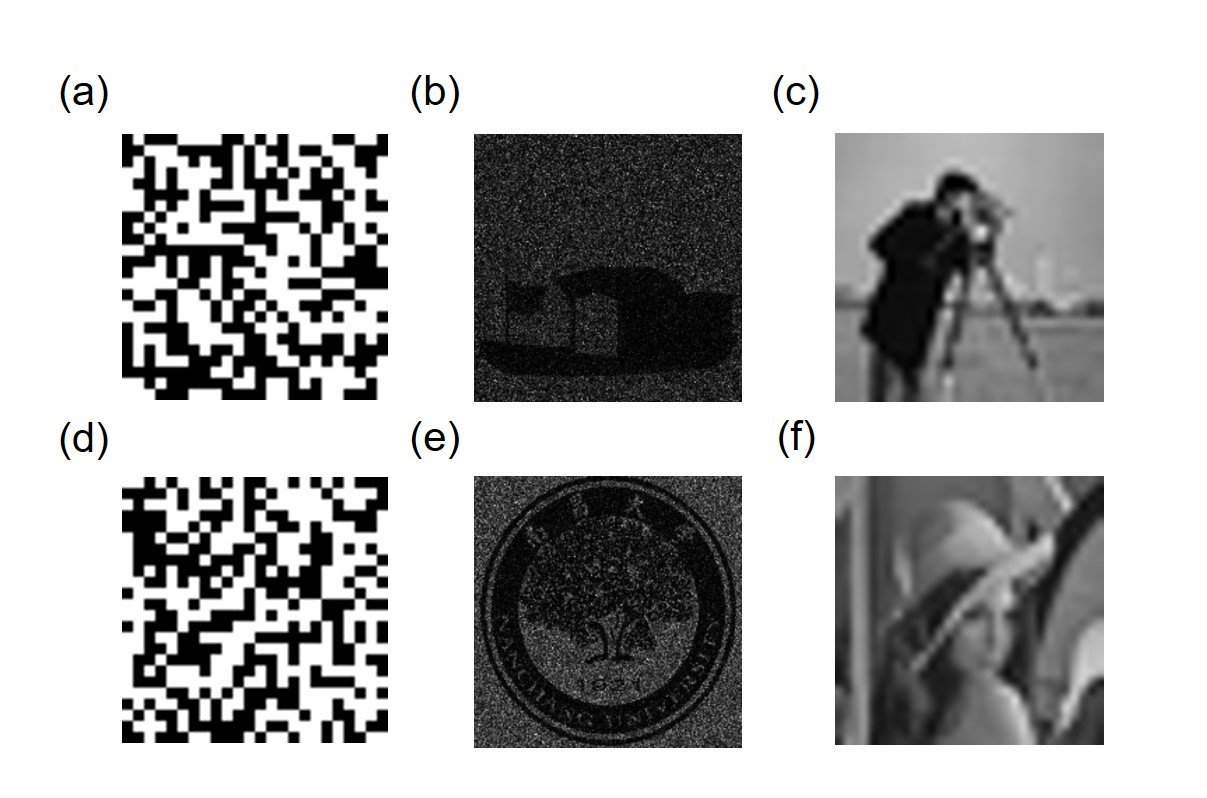}
	\caption{Simulation results for the dual near-field and dual far-field channels. (a), (b) Under \textit{x}-polarized illumination, near-field and far-field images. (d), (e) Under \textit{y}-polarized illumination, near-field and far-field images. (c), (f) decrypted reconstructions from the near-field QR images.}
	\label{fig4}
\end{figure*}
We then perform critical post-processing: each pixel's grayscale value is expanded into a binary sequence with a reserved termination bit, then reorganized into a 3×3 matrix placed at the original pixel position. This transformation converts continuous grayscale ciphertext into binary QR code images while simultaneously embedding essential key information. Crucially, the conversion from continuous grayscale to binary QR patterns significantly relaxes the metasurface's imaging precision requirements, as binary representation inherently tolerates optical aberrations and imaging inaccuracies that would degrade higher-fidelity grayscale reconstruction. This adaptation effectively bridges chaotic encryption with metasurface-based optical imaging, enhancing both security and robustness in the system. Moreover, embedding decryption keys directly within the QR-encoded ciphertext creates a self-contained cryptographic unit. This eliminates the conventional risks associated with separate key transmission, guarantees immediate key availability during optical decryption, and effectively mitigates vulnerabilities inherent in traditional key distribution channels.

\begin{figure*}[htbp]
	\centering
	\includegraphics[width=0.9\textwidth]{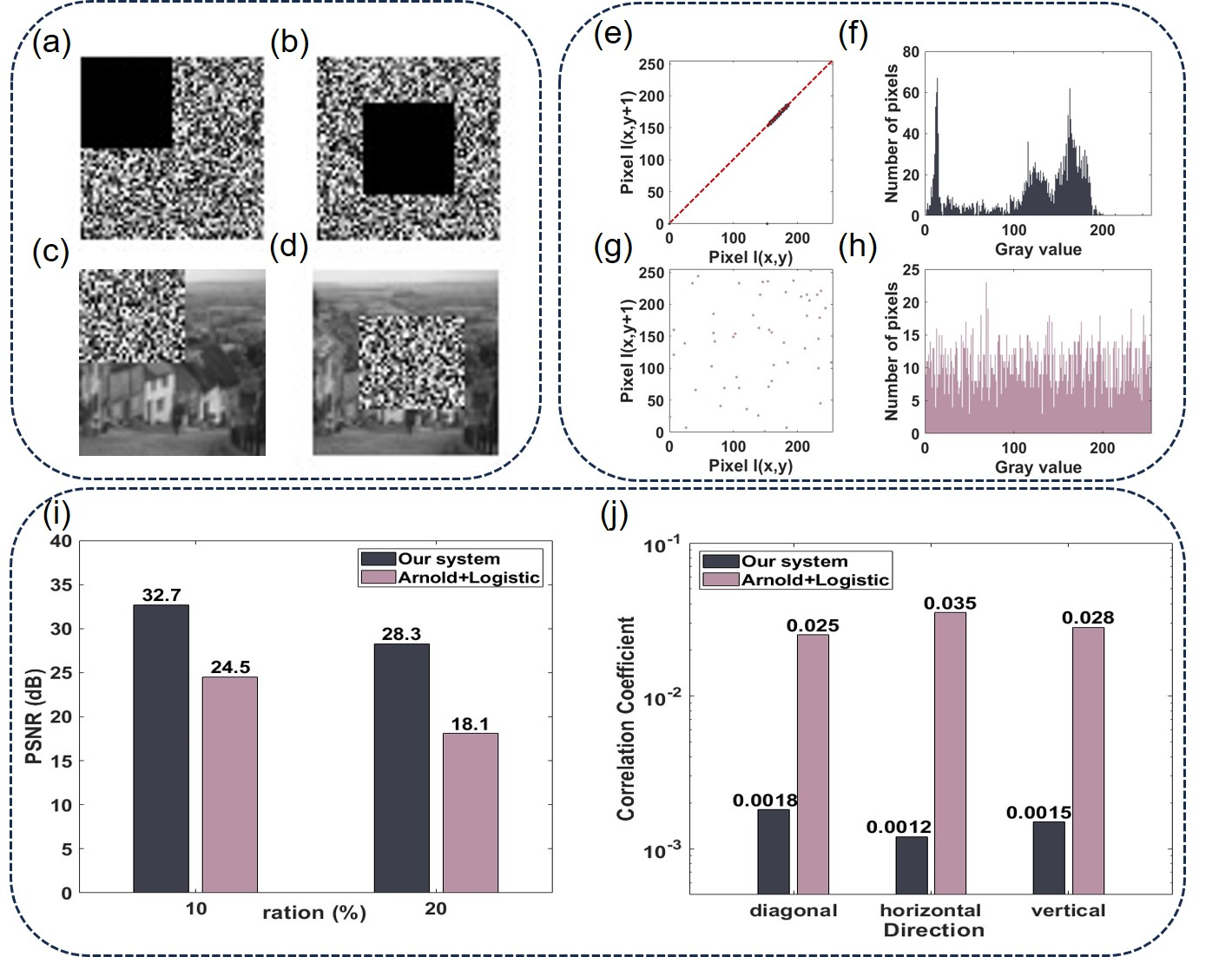}
	\caption{Chaos encryption performance evaluation. (a), (b) Cropping attacks performed on the encrypted QR code images. (c), (d) Decryption results from the cropped ciphertext. (e), (g) The adjacent pixel correlation analysis before and after encryption respectively. (f), (h) Grayscale histograms before and after encryption respectively. (i), (j) Comparative performance against Arnold encryption, including signal-to-noise ratio under cropping attacks and three-dimensional adjacent pixel correlation measurements across horizontal, vertical and diagonal directions.} 
	\label{fig5}
\end{figure*}
This encryption framework establishes a robust security architecture leveraging chaotic sequences and DNA encoding. The inherent sensitivity of chaotic systems to initial conditions, combined with dynamic switching of DNA encoding rules, effectively disrupts statistical correlations and prevents pattern-fixed attacks. This combined mechanism provides a comprehensive defense against statistical analysis, cropping attempts, and brute-force attacks. Notably, the innovative QR code reprocessing technique addresses two fundamental challenges: firstly, it significantly reduces the precision requirements for chaotic encryption imaging by converting continuous grayscale to binary representation, thereby effectively mitigating optical decryption errors induced by metasurface limitations; secondly, it seamlessly integrates key information within the encrypted patterns, overcoming traditional key distribution vulnerabilities while maintaining system complexity. This synergistic enhancement substantially strengthen the overall security framework.

\subsection{Imaging results and encryption performance tests}

By precise arrangement of $\mathrm{Sb_2}$$\mathrm{S_3}$ nanobricks on a single metasurface, we demonstrate two near-field QR code imaging channels and concurrently implement two far-field holographic channels to fully exploit the device's information capacity. Simulation results in Fig. 4 verify this multifunctional operation. Under $x$-polarized illumination, Figs. 4(a) and 4(b) present the near-field nanoprinting and the far-field holographic imaging, respectively. Correspondingly, Figs. 4(d) and 4(e) present the other near-field nanoprinting channel and far-field hologram under $y$-polarized illumination. All four channels exhibit high fidelity. Both near-field nanoprinting channels exhibit a Structural Similarity Index (SSIM) of 99\%, while the far-field holograms reach SSIM = 85.43\% for the NCU emblem and SSIM = 72.04\% for the boat.

The system employs a dual-layer encryption architecture that combines algorithmic security with physical protection through the reversible state transition of Sb$_2$S$_3$. This mechanism ensures optical concealment of the ciphertext images throughout the crystalline state of Sb$_2$S$_3$. Decryption is enabled exclusively upon switching to the amorphous state, thus establishing a dual-layer security mechanism. Figs. 4(c) and 4(f) show decrypted reconstruction results obtained directly from the near-field QR images after physical decryption, confirming essentially complete preservation of the original plaintext. The residual blur in the decrypted images is attributable to the finite metasurface size (plaintext resolution fixed at 32×32 pixels) rather than to intrinsic imaging degradation, as evidenced by higher-resolution decryptions obtained from larger metasurfaces provided in the Supplementary Material. Additional results illustrating the physical encryption concealment effect are also included in the Supplementary Material.

To quantitatively evaluate the performance of the chaotic-encryption scheme, we carry out a series of robustness and statistical tests. We first evaluate the scheme against cropping attacks. This analysis serves two purposes: (i) to quantify resilience to data loss during transmission or storage by determining if partial ciphertexts retain information confidentiality, and (ii) to determine whether localized errors in the ciphertext propagate into the reconstruction and consequently degrade the global decryption fidelity. Figs. 5(a) and 5(b) illustrate severe data loss scenarios through quarter-region removal at image peripheries and centers, respectively. The corresponding decrypted outputs in Figs. 5(c) and 5(d) recover all information outside the masked regions. These results demonstrate scheme's resistance to cropping attacks, that discarding 25\% of ciphertext—whether from edges or central areas—enables complete plaintext recovery outside affected zones. This resistance to localized data loss preserves confidentiality and integrity during transmission or storage and prevents local ciphertext corruption from propagating into the global reconstruction, thereby markedly enhancing the scheme's practical robustness.

Next, we analyze the pixel gray histogram and the correlation coefficients to evaluate the encryption effectiveness from statistical and structural perspectives. The histogram reflects the distribution of pixel intensities, where a uniform distribution indicates resistance to statistical attacks, while the pixel correlation coefficient quantifies the gray-level similarity between adjacent pixels, with lower correlation indicating stronger encryption. Fig. 5(e) reveals significant linear clustering in horizontal adjacent-pixel correlations for the plaintext, characteristic of natural images. In contrast, the same analysis for the ciphertext in Fig. 5(g) shows a nearly uniform dispersion, confirming effective spatial decorrelation introduced by the chaotic transformation. Complementary histograms in Figs. 5(f) and 5(h) demonstrate encryption-induced flattening of the statistically distinctive plaintext gray distribution into an approximately uniform profile, thereby reducing vulnerability to histogram-based attacks. Considering that the adopted metasurface size with 32×32 pixels limits plaintext resolution, larger-scale correlation and histogram comparisons are provided in the Supplementary Material. Finally, to benchmark our approach we compare it with a conventional Arnold–Logistic scheme. As shown in Figs. 5(i) and 5(j), our method achieves superior signal-to-noise ratios of 32.7 dB and 28.3 dB under 10\% and 20\% cropping attacks, respectively, significantly outperforming the Arnold-Logistic system, which yields 24.5 dB and 18.1 dB under the same conditions. Moreover, the adjacent-pixel correlation coefficients of our system are substantially lower across horizontal, vertical, and diagonal directions, measuring 0.0012, 0.0015, and 0.0018, respectively, compared to 0.035, 0.028, and 0.025 for the Arnold-Logistic approach. These results demonstrate the enhanced robustness and statistical security of our encryption system against both data loss and statistical attacks, which are critical for reliable and secure image transmission. These results validate that the proposed chaotic encryption achieves both strong statistical concealment and practical robustness against localized data loss.

\section{Conclusions}

In this work, we have demonstrated a high-security optical encryption platform that synergistically integrates a hyperchaotic encryption algorithm with a dynamically reconfigurable phase-change metasurface. By leveraging polarization multiplexing and double-cell meta-atom design, the metasurface simultaneously supports two near-field nanoprinting channels and two far-field holographic channels, achieving high-fidelity imaging with SSIM values up to 99\% for near-field QR codes and with values ranging from 72.04\% to 85.43\% for far-field holograms. The incorporation of Chen hyperchaotic encryption combined with Logistic mapping and dynamic DNA encoding ensures robust cryptographic security, significantly enhancing resistance against statistical and cropping attacks. Moreover, the phase-change functionality of Sb$_2$S$_3$ enables physical-layer security through reversible crystalline-to-amorphous transitions, allowing dynamic concealment of ciphertext images and adding an unclonable physical dimension to the encryption system. This dual-layer algorithm–physical co-security framework provides a compact, high-capacity, and reliable solution for applications in secure optical communication and multi-dimensional data storage, paving the way for next-generation optical information encryption technologies.

\begin{acknowledgments}	
	
This work was supported by the National Natural Science Foundation of China (Grants No. 12364045, No. 12264028, and No. 12304420), the Natural Science Foundation of Jiangxi Province (Grants No. 20232BAB201040 and No. 20232BAB211025), and the Young Elite Scientists Sponsorship Program by JXAST (Grants No. 2023QT11 and No. 2025QT04). 

\end{acknowledgments}


\begin{thebibliography}{64}%
	\makeatletter
	\providecommand \@ifxundefined [1]{%
		\@ifx{#1\undefined}
	}%
	\providecommand \@ifnum [1]{%
		\ifnum #1\expandafter \@firstoftwo
		\else \expandafter \@secondoftwo
		\fi
	}%
	\providecommand \@ifx [1]{%
		\ifx #1\expandafter \@firstoftwo
		\else \expandafter \@secondoftwo
		\fi
	}%
	\providecommand \natexlab [1]{#1}%
	\providecommand \enquote  [1]{``#1''}%
	\providecommand \bibnamefont  [1]{#1}%
	\providecommand \bibfnamefont [1]{#1}%
	\providecommand \citenamefont [1]{#1}%
	\providecommand \href@noop [0]{\@secondoftwo}%
	\providecommand \href [0]{\begingroup \@sanitize@url \@href}%
	\providecommand \@href[1]{\@@startlink{#1}\@@href}%
	\providecommand \@@href[1]{\endgroup#1\@@endlink}%
	\providecommand \@sanitize@url [0]{\catcode `\\12\catcode `\$12\catcode
		`\&12\catcode `\#12\catcode `\^12\catcode `\_12\catcode `\%12\relax}%
	\providecommand \@@startlink[1]{}%
	\providecommand \@@endlink[0]{}%
	\providecommand \url  [0]{\begingroup\@sanitize@url \@url }%
	\providecommand \@url [1]{\endgroup\@href {#1}{\urlprefix }}%
	\providecommand \urlprefix  [0]{URL }%
	\providecommand \Eprint [0]{\href }%
	\providecommand \doibase [0]{https://doi.org/}%
	\providecommand \selectlanguage [0]{\@gobble}%
	\providecommand \bibinfo  [0]{\@secondoftwo}%
	\providecommand \bibfield  [0]{\@secondoftwo}%
	\providecommand \translation [1]{[#1]}%
	\providecommand \BibitemOpen [0]{}%
	\providecommand \bibitemStop [0]{}%
	\providecommand \bibitemNoStop [0]{.\EOS\space}%
	\providecommand \EOS [0]{\spacefactor3000\relax}%
	\providecommand \BibitemShut  [1]{\csname bibitem#1\endcsname}%
	\let\auto@bib@innerbib\@empty
	\bibitem [{\citenamefont {Xu}\ \emph {et~al.}(2025{\natexlab{a}})\citenamefont
		{Xu}, \citenamefont {Wei}, \citenamefont {Zhang}, \citenamefont {Galdi},
		\citenamefont {Li},\ and\ \citenamefont {Cui}}]{xu2025chaotic}%
	\BibitemOpen
	\bibfield  {author} {\bibinfo {author} {\bibfnamefont {J.~W.}\ \bibnamefont
			{Xu}}, \bibinfo {author} {\bibfnamefont {M.}~\bibnamefont {Wei}}, \bibinfo
		{author} {\bibfnamefont {L.}~\bibnamefont {Zhang}}, \bibinfo {author}
		{\bibfnamefont {V.}~\bibnamefont {Galdi}}, \bibinfo {author} {\bibfnamefont
			{L.}~\bibnamefont {Li}},\ and\ \bibinfo {author} {\bibfnamefont {T.~J.}\
			\bibnamefont {Cui}},\ }\bibfield  {title} {\bibinfo {title} {Chaotic
			information metasurface for direct physical-layer secure communication},\
	}\href@noop {} {\bibfield  {journal} {\bibinfo  {journal} {Nat. Commun.}\
		}\textbf {\bibinfo {volume} {16}},\ \bibinfo {pages} {5853} (\bibinfo {year}
		{2025}{\natexlab{a}})}\BibitemShut {NoStop}%
	\bibitem [{\citenamefont {Ning}\ \emph {et~al.}(2025)\citenamefont {Ning},
		\citenamefont {Zhong}, \citenamefont {Gu}, \citenamefont {Zhang},
		\citenamefont {Qu}, \citenamefont {Ding}, \citenamefont {Li},\ and\
		\citenamefont {Li}}]{ning2025enhanced}%
	\BibitemOpen
	\bibfield  {author} {\bibinfo {author} {\bibfnamefont {M.}~\bibnamefont
			{Ning}}, \bibinfo {author} {\bibfnamefont {H.}~\bibnamefont {Zhong}},
		\bibinfo {author} {\bibfnamefont {Z.}~\bibnamefont {Gu}}, \bibinfo {author}
		{\bibfnamefont {L.-E.}\ \bibnamefont {Zhang}}, \bibinfo {author}
		{\bibfnamefont {N.}~\bibnamefont {Qu}}, \bibinfo {author} {\bibfnamefont
			{J.}~\bibnamefont {Ding}}, \bibinfo {author} {\bibfnamefont {T.}~\bibnamefont
			{Li}},\ and\ \bibinfo {author} {\bibfnamefont {L.}~\bibnamefont {Li}},\
	}\bibfield  {title} {\bibinfo {title} {Enhanced optical encryption via
			polarization-dependent multi-channel metasurfaces},\ }\href@noop {}
	{\bibfield  {journal} {\bibinfo  {journal} {Nanophotonics}\ }\textbf
		{\bibinfo {volume} {14}},\ \bibinfo {pages} {495} (\bibinfo {year}
		{2025})}\BibitemShut {NoStop}%
	\bibitem [{\citenamefont {Ouyang}\ \emph {et~al.}(2021)\citenamefont {Ouyang},
		\citenamefont {Yu}, \citenamefont {Liu}, \citenamefont {Ma},\ and\
		\citenamefont {Yu}}]{ouyang2021underwater}%
	\BibitemOpen
	\bibfield  {author} {\bibinfo {author} {\bibfnamefont {F.}~\bibnamefont
			{Ouyang}}, \bibinfo {author} {\bibfnamefont {J.}~\bibnamefont {Yu}}, \bibinfo
		{author} {\bibfnamefont {H.}~\bibnamefont {Liu}}, \bibinfo {author}
		{\bibfnamefont {Z.}~\bibnamefont {Ma}},\ and\ \bibinfo {author}
		{\bibfnamefont {X.}~\bibnamefont {Yu}},\ }\bibfield  {title} {\bibinfo
		{title} {Underwater imaging system based on light field technology},\
	}\href@noop {} {\bibfield  {journal} {\bibinfo  {journal} {IEEE Sensors J.}\
		}\textbf {\bibinfo {volume} {21}},\ \bibinfo {pages} {13753} (\bibinfo {year}
		{2021})}\BibitemShut {NoStop}%
	\bibitem [{\citenamefont {Feng}\ \emph {et~al.}(2023)\citenamefont {Feng},
		\citenamefont {He}, \citenamefont {Shi}, \citenamefont {Song}, \citenamefont
		{Zhu}, \citenamefont {Zhang}, \citenamefont {Wang}, \citenamefont {Tsai},\
		and\ \citenamefont {Cheng}}]{feng2023diatomic}%
	\BibitemOpen
	\bibfield  {author} {\bibinfo {author} {\bibfnamefont {C.}~\bibnamefont
			{Feng}}, \bibinfo {author} {\bibfnamefont {T.}~\bibnamefont {He}}, \bibinfo
		{author} {\bibfnamefont {Y.}~\bibnamefont {Shi}}, \bibinfo {author}
		{\bibfnamefont {Q.}~\bibnamefont {Song}}, \bibinfo {author} {\bibfnamefont
			{J.}~\bibnamefont {Zhu}}, \bibinfo {author} {\bibfnamefont {J.}~\bibnamefont
			{Zhang}}, \bibinfo {author} {\bibfnamefont {Z.}~\bibnamefont {Wang}},
		\bibinfo {author} {\bibfnamefont {D.~P.}\ \bibnamefont {Tsai}},\ and\
		\bibinfo {author} {\bibfnamefont {X.}~\bibnamefont {Cheng}},\ }\bibfield
	{title} {\bibinfo {title} {Diatomic metasurface for efficient six-channel
			modulation of jones matrix (laser photonics rev. 17 (8)/2023)},\ }\href@noop
	{} {\bibfield  {journal} {\bibinfo  {journal} {Laser Photonics Rev.}\
		}\textbf {\bibinfo {volume} {17}},\ \bibinfo {pages} {2370040} (\bibinfo
		{year} {2023})}\BibitemShut {NoStop}%
	\bibitem [{\citenamefont {Yin}\ \emph {et~al.}(2024)\citenamefont {Yin},
		\citenamefont {Jiang}, \citenamefont {Wang}, \citenamefont {Liu},
		\citenamefont {Xie}, \citenamefont {Wang}, \citenamefont {Wang},\ and\
		\citenamefont {Huang}}]{yin2024multi}%
	\BibitemOpen
	\bibfield  {author} {\bibinfo {author} {\bibfnamefont {Y.}~\bibnamefont
			{Yin}}, \bibinfo {author} {\bibfnamefont {Q.}~\bibnamefont {Jiang}}, \bibinfo
		{author} {\bibfnamefont {H.}~\bibnamefont {Wang}}, \bibinfo {author}
		{\bibfnamefont {J.}~\bibnamefont {Liu}}, \bibinfo {author} {\bibfnamefont
			{Y.}~\bibnamefont {Xie}}, \bibinfo {author} {\bibfnamefont {Q.}~\bibnamefont
			{Wang}}, \bibinfo {author} {\bibfnamefont {Y.}~\bibnamefont {Wang}},\ and\
		\bibinfo {author} {\bibfnamefont {L.}~\bibnamefont {Huang}},\ }\bibfield
	{title} {\bibinfo {title} {Multi-dimensional multiplexed metasurface
			holography by inverse design},\ }\href@noop {} {\bibfield  {journal}
		{\bibinfo  {journal} {Adv. Mater.}\ }\textbf {\bibinfo {volume} {36}},\
		\bibinfo {pages} {2312303} (\bibinfo {year} {2024})}\BibitemShut {NoStop}%
	\bibitem [{\citenamefont {Sun}\ \emph {et~al.}(2012)\citenamefont {Sun},
		\citenamefont {He}, \citenamefont {Xiao}, \citenamefont {Xu}, \citenamefont
		{Li},\ and\ \citenamefont {Zhou}}]{sun2012gradient}%
	\BibitemOpen
	\bibfield  {author} {\bibinfo {author} {\bibfnamefont {S.}~\bibnamefont
			{Sun}}, \bibinfo {author} {\bibfnamefont {Q.}~\bibnamefont {He}}, \bibinfo
		{author} {\bibfnamefont {S.}~\bibnamefont {Xiao}}, \bibinfo {author}
		{\bibfnamefont {Q.}~\bibnamefont {Xu}}, \bibinfo {author} {\bibfnamefont
			{X.}~\bibnamefont {Li}},\ and\ \bibinfo {author} {\bibfnamefont
			{L.}~\bibnamefont {Zhou}},\ }\bibfield  {title} {\bibinfo {title}
		{Gradient-index meta-surfaces as a bridge linking propagating waves and
			surface waves},\ }\href@noop {} {\bibfield  {journal} {\bibinfo  {journal}
			{Nat. Mater.}\ }\textbf {\bibinfo {volume} {11}},\ \bibinfo {pages} {426}
		(\bibinfo {year} {2012})}\BibitemShut {NoStop}%
	\bibitem [{\citenamefont {Li}\ \emph {et~al.}(2024)\citenamefont {Li},
		\citenamefont {Yu}, \citenamefont {Qiu},\ and\ \citenamefont
		{Qi}}]{li2024intelligent}%
	\BibitemOpen
	\bibfield  {author} {\bibinfo {author} {\bibfnamefont {W.}~\bibnamefont
			{Li}}, \bibinfo {author} {\bibfnamefont {Q.}~\bibnamefont {Yu}}, \bibinfo
		{author} {\bibfnamefont {J.~H.}\ \bibnamefont {Qiu}},\ and\ \bibinfo {author}
		{\bibfnamefont {J.}~\bibnamefont {Qi}},\ }\bibfield  {title} {\bibinfo
		{title} {Intelligent wireless power transfer via a 2-bit compact
			reconfigurable transmissive-metasurface-based router},\ }\href@noop {}
	{\bibfield  {journal} {\bibinfo  {journal} {Nat. Commun.}\ }\textbf {\bibinfo
			{volume} {15}},\ \bibinfo {pages} {2807} (\bibinfo {year}
		{2024})}\BibitemShut {NoStop}%
	\bibitem [{\citenamefont {Yu}\ \emph {et~al.}(2011)\citenamefont {Yu},
		\citenamefont {Genevet}, \citenamefont {Kats}, \citenamefont {Aieta},
		\citenamefont {Tetienne}, \citenamefont {Capasso},\ and\ \citenamefont
		{Gaburro}}]{yu2011light}%
	\BibitemOpen
	\bibfield  {author} {\bibinfo {author} {\bibfnamefont {N.}~\bibnamefont
			{Yu}}, \bibinfo {author} {\bibfnamefont {P.}~\bibnamefont {Genevet}},
		\bibinfo {author} {\bibfnamefont {M.~A.}\ \bibnamefont {Kats}}, \bibinfo
		{author} {\bibfnamefont {F.}~\bibnamefont {Aieta}}, \bibinfo {author}
		{\bibfnamefont {J.-P.}\ \bibnamefont {Tetienne}}, \bibinfo {author}
		{\bibfnamefont {F.}~\bibnamefont {Capasso}},\ and\ \bibinfo {author}
		{\bibfnamefont {Z.}~\bibnamefont {Gaburro}},\ }\bibfield  {title} {\bibinfo
		{title} {Light propagation with phase discontinuities: generalized laws of
			reflection and refraction},\ }\href@noop {} {\bibfield  {journal} {\bibinfo
			{journal} {Science}\ }\textbf {\bibinfo {volume} {334}},\ \bibinfo {pages}
		{333} (\bibinfo {year} {2011})}\BibitemShut {NoStop}%
	\bibitem [{\citenamefont {Frese}\ \emph {et~al.}(2019)\citenamefont {Frese},
		\citenamefont {Wei}, \citenamefont {Wang}, \citenamefont {Huang},\ and\
		\citenamefont {Zentgraf}}]{frese2019nonreciprocal}%
	\BibitemOpen
	\bibfield  {author} {\bibinfo {author} {\bibfnamefont {D.}~\bibnamefont
			{Frese}}, \bibinfo {author} {\bibfnamefont {Q.}~\bibnamefont {Wei}}, \bibinfo
		{author} {\bibfnamefont {Y.}~\bibnamefont {Wang}}, \bibinfo {author}
		{\bibfnamefont {L.}~\bibnamefont {Huang}},\ and\ \bibinfo {author}
		{\bibfnamefont {T.}~\bibnamefont {Zentgraf}},\ }\bibfield  {title} {\bibinfo
		{title} {Nonreciprocal asymmetric polarization encryption by layered
			plasmonic metasurfaces},\ }\href@noop {} {\bibfield  {journal} {\bibinfo
			{journal} {Nano Lett.}\ }\textbf {\bibinfo {volume} {19}},\ \bibinfo {pages}
		{3976} (\bibinfo {year} {2019})}\BibitemShut {NoStop}%
	\bibitem [{\citenamefont {Deng}\ \emph
		{et~al.}(2020{\natexlab{a}})\citenamefont {Deng}, \citenamefont {Deng},
		\citenamefont {Guan}, \citenamefont {Tao}, \citenamefont {Chen},
		\citenamefont {Yang}, \citenamefont {Zhang}, \citenamefont {Tang},
		\citenamefont {Li}, \citenamefont {Li} \emph {et~al.}}]{deng2020malus}%
	\BibitemOpen
	\bibfield  {author} {\bibinfo {author} {\bibfnamefont {L.}~\bibnamefont
			{Deng}}, \bibinfo {author} {\bibfnamefont {J.}~\bibnamefont {Deng}}, \bibinfo
		{author} {\bibfnamefont {Z.}~\bibnamefont {Guan}}, \bibinfo {author}
		{\bibfnamefont {J.}~\bibnamefont {Tao}}, \bibinfo {author} {\bibfnamefont
			{Y.}~\bibnamefont {Chen}}, \bibinfo {author} {\bibfnamefont {Y.}~\bibnamefont
			{Yang}}, \bibinfo {author} {\bibfnamefont {D.}~\bibnamefont {Zhang}},
		\bibinfo {author} {\bibfnamefont {J.}~\bibnamefont {Tang}}, \bibinfo {author}
		{\bibfnamefont {Z.}~\bibnamefont {Li}}, \bibinfo {author} {\bibfnamefont
			{Z.}~\bibnamefont {Li}}, \emph {et~al.},\ }\bibfield  {title} {\bibinfo
		{title} {Malus-metasurface-assisted polarization multiplexing},\ }\href@noop
	{} {\bibfield  {journal} {\bibinfo  {journal} {Light Sci. Appl.}\ }\textbf
		{\bibinfo {volume} {9}},\ \bibinfo {pages} {101} (\bibinfo {year}
		{2020}{\natexlab{a}})}\BibitemShut {NoStop}%
	\bibitem [{\citenamefont {Xu}\ \emph {et~al.}(2025{\natexlab{b}})\citenamefont
		{Xu}, \citenamefont {Li}, \citenamefont {Liu}, \citenamefont {Li},
		\citenamefont {Tan}, \citenamefont {Xu}, \citenamefont {Zhang},\ and\
		\citenamefont {Yao}}]{Xu2025}%
	\BibitemOpen
	\bibfield  {author} {\bibinfo {author} {\bibfnamefont {W.}~\bibnamefont
			{Xu}}, \bibinfo {author} {\bibfnamefont {H.}~\bibnamefont {Li}}, \bibinfo
		{author} {\bibfnamefont {Y.}~\bibnamefont {Liu}}, \bibinfo {author}
		{\bibfnamefont {J.}~\bibnamefont {Li}}, \bibinfo {author} {\bibfnamefont
			{Q.}~\bibnamefont {Tan}}, \bibinfo {author} {\bibfnamefont {H.}~\bibnamefont
			{Xu}}, \bibinfo {author} {\bibfnamefont {Y.}~\bibnamefont {Zhang}},\ and\
		\bibinfo {author} {\bibfnamefont {J.}~\bibnamefont {Yao}},\ }\bibfield
	{title} {\bibinfo {title} {Generalized metasurface design for polarization
			reconstruction in a single snapshot},\ }\href
	{https://doi.org/10.1109/jlt.2025.3565909} {\bibfield  {journal} {\bibinfo
			{journal} {J. Lightwave Technol.}\ }\textbf {\bibinfo {volume} {43}},\
		\bibinfo {pages} {6807} (\bibinfo {year} {2025}{\natexlab{b}})}\BibitemShut
	{NoStop}%
	\bibitem [{\citenamefont {Liu}\ \emph {et~al.}(2021)\citenamefont {Liu},
		\citenamefont {Zhu}, \citenamefont {Huo}, \citenamefont {Feng}, \citenamefont
		{Song}, \citenamefont {Zhang}, \citenamefont {Chen}, \citenamefont {Lezec},
		\citenamefont {Lu}, \citenamefont {Agrawal} \emph
		{et~al.}}]{liu2021multifunctional}%
	\BibitemOpen
	\bibfield  {author} {\bibinfo {author} {\bibfnamefont {M.}~\bibnamefont
			{Liu}}, \bibinfo {author} {\bibfnamefont {W.}~\bibnamefont {Zhu}}, \bibinfo
		{author} {\bibfnamefont {P.}~\bibnamefont {Huo}}, \bibinfo {author}
		{\bibfnamefont {L.}~\bibnamefont {Feng}}, \bibinfo {author} {\bibfnamefont
			{M.}~\bibnamefont {Song}}, \bibinfo {author} {\bibfnamefont {C.}~\bibnamefont
			{Zhang}}, \bibinfo {author} {\bibfnamefont {L.}~\bibnamefont {Chen}},
		\bibinfo {author} {\bibfnamefont {H.~J.}\ \bibnamefont {Lezec}}, \bibinfo
		{author} {\bibfnamefont {Y.}~\bibnamefont {Lu}}, \bibinfo {author}
		{\bibfnamefont {A.}~\bibnamefont {Agrawal}}, \emph {et~al.},\ }\bibfield
	{title} {\bibinfo {title} {Multifunctional metasurfaces enabled by
			simultaneous and independent control of phase and amplitude for orthogonal
			polarization states},\ }\href@noop {} {\bibfield  {journal} {\bibinfo
			{journal} {Light Sci. Appl.}\ }\textbf {\bibinfo {volume} {10}},\ \bibinfo
		{pages} {107} (\bibinfo {year} {2021})}\BibitemShut {NoStop}%
	\bibitem [{\citenamefont {Lu}\ \emph {et~al.}(2024)\citenamefont {Lu},
		\citenamefont {Qiu}, \citenamefont {Liu}, \citenamefont {Zhang},
		\citenamefont {Xiao},\ and\ \citenamefont {Yu}}]{lu2024metasurface}%
	\BibitemOpen
	\bibfield  {author} {\bibinfo {author} {\bibfnamefont {G.}~\bibnamefont
			{Lu}}, \bibinfo {author} {\bibfnamefont {J.}~\bibnamefont {Qiu}}, \bibinfo
		{author} {\bibfnamefont {T.}~\bibnamefont {Liu}}, \bibinfo {author}
		{\bibfnamefont {D.}~\bibnamefont {Zhang}}, \bibinfo {author} {\bibfnamefont
			{S.}~\bibnamefont {Xiao}},\ and\ \bibinfo {author} {\bibfnamefont
			{T.}~\bibnamefont {Yu}},\ }\bibfield  {title} {\bibinfo {title}
		{Metasurface-based diffractive optical networks with dual-channel complex
			amplitude modulation},\ }\href@noop {} {\bibfield  {journal} {\bibinfo
			{journal} {J. Lightwave Technol.}\ }\textbf {\bibinfo {volume} {42}},\
		\bibinfo {pages} {7282} (\bibinfo {year} {2024})}\BibitemShut {NoStop}%
	\bibitem [{\citenamefont {Xie}\ \emph {et~al.}(2021)\citenamefont {Xie},
		\citenamefont {Pu}, \citenamefont {Jin}, \citenamefont {Xu}, \citenamefont
		{Guo}, \citenamefont {Li}, \citenamefont {Gao}, \citenamefont {Ma},\ and\
		\citenamefont {Luo}}]{xie2021generalized}%
	\BibitemOpen
	\bibfield  {author} {\bibinfo {author} {\bibfnamefont {X.}~\bibnamefont
			{Xie}}, \bibinfo {author} {\bibfnamefont {M.}~\bibnamefont {Pu}}, \bibinfo
		{author} {\bibfnamefont {J.}~\bibnamefont {Jin}}, \bibinfo {author}
		{\bibfnamefont {M.}~\bibnamefont {Xu}}, \bibinfo {author} {\bibfnamefont
			{Y.}~\bibnamefont {Guo}}, \bibinfo {author} {\bibfnamefont {X.}~\bibnamefont
			{Li}}, \bibinfo {author} {\bibfnamefont {P.}~\bibnamefont {Gao}}, \bibinfo
		{author} {\bibfnamefont {X.}~\bibnamefont {Ma}},\ and\ \bibinfo {author}
		{\bibfnamefont {X.}~\bibnamefont {Luo}},\ }\bibfield  {title} {\bibinfo
		{title} {Generalized pancharatnam-berry phase in rotationally symmetric
			meta-atoms},\ }\href@noop {} {\bibfield  {journal} {\bibinfo  {journal}
			{Phys. Rev. Lett.}\ }\textbf {\bibinfo {volume} {126}},\ \bibinfo {pages}
		{183902} (\bibinfo {year} {2021})}\BibitemShut {NoStop}%
	\bibitem [{\citenamefont {Li}\ \emph {et~al.}(2025{\natexlab{a}})\citenamefont
		{Li}, \citenamefont {Nan}, \citenamefont {Tian}, \citenamefont {Lv},
		\citenamefont {Xu}, \citenamefont {Tan}, \citenamefont {Li}, \citenamefont
		{Li}, \citenamefont {Luo}, \citenamefont {Tang}, \citenamefont {Zhang},\ and\
		\citenamefont {Yao}}]{Li2025}%
	\BibitemOpen
	\bibfield  {author} {\bibinfo {author} {\bibfnamefont {J.}~\bibnamefont
			{Li}}, \bibinfo {author} {\bibfnamefont {T.}~\bibnamefont {Nan}}, \bibinfo
		{author} {\bibfnamefont {H.}~\bibnamefont {Tian}}, \bibinfo {author}
		{\bibfnamefont {Y.}~\bibnamefont {Lv}}, \bibinfo {author} {\bibfnamefont
			{H.}~\bibnamefont {Xu}}, \bibinfo {author} {\bibfnamefont {Q.}~\bibnamefont
			{Tan}}, \bibinfo {author} {\bibfnamefont {H.}~\bibnamefont {Li}}, \bibinfo
		{author} {\bibfnamefont {J.}~\bibnamefont {Li}}, \bibinfo {author}
		{\bibfnamefont {L.}~\bibnamefont {Luo}}, \bibinfo {author} {\bibfnamefont
			{T.}~\bibnamefont {Tang}}, \bibinfo {author} {\bibfnamefont {Y.}~\bibnamefont
			{Zhang}},\ and\ \bibinfo {author} {\bibfnamefont {J.}~\bibnamefont {Yao}},\
	}\bibfield  {title} {\bibinfo {title} {Extremely simplified binary-phase
			metasurfaces for circularly polarized terahertz waves manipulation},\ }\href
	{https://doi.org/10.1109/jlt.2024.3515058} {\bibfield  {journal} {\bibinfo
			{journal} {J. Lightwave Technol.}\ }\textbf {\bibinfo {volume} {43}},\
		\bibinfo {pages} {3413} (\bibinfo {year} {2025}{\natexlab{a}})}\BibitemShut
	{NoStop}%
	\bibitem [{\citenamefont {Zhang}\ \emph
		{et~al.}(2019{\natexlab{a}})\citenamefont {Zhang}, \citenamefont {Pu},
		\citenamefont {Guo}, \citenamefont {Jin}, \citenamefont {Li}, \citenamefont
		{Ma}, \citenamefont {Luo}, \citenamefont {Wang},\ and\ \citenamefont
		{Luo}}]{zhang2019colorful}%
	\BibitemOpen
	\bibfield  {author} {\bibinfo {author} {\bibfnamefont {X.}~\bibnamefont
			{Zhang}}, \bibinfo {author} {\bibfnamefont {M.}~\bibnamefont {Pu}}, \bibinfo
		{author} {\bibfnamefont {Y.}~\bibnamefont {Guo}}, \bibinfo {author}
		{\bibfnamefont {J.}~\bibnamefont {Jin}}, \bibinfo {author} {\bibfnamefont
			{X.}~\bibnamefont {Li}}, \bibinfo {author} {\bibfnamefont {X.}~\bibnamefont
			{Ma}}, \bibinfo {author} {\bibfnamefont {J.}~\bibnamefont {Luo}}, \bibinfo
		{author} {\bibfnamefont {C.}~\bibnamefont {Wang}},\ and\ \bibinfo {author}
		{\bibfnamefont {X.}~\bibnamefont {Luo}},\ }\bibfield  {title} {\bibinfo
		{title} {Colorful metahologram with independently controlled images in
			transmission and reflection spaces},\ }\href@noop {} {\bibfield  {journal}
		{\bibinfo  {journal} {Adv. Funct. Mater.}\ }\textbf {\bibinfo {volume}
			{29}},\ \bibinfo {pages} {1809145} (\bibinfo {year}
		{2019}{\natexlab{a}})}\BibitemShut {NoStop}%
	\bibitem [{\citenamefont {Wang}\ \emph {et~al.}(2016)\citenamefont {Wang},
		\citenamefont {Dong}, \citenamefont {Li}, \citenamefont {Yang}, \citenamefont
		{Sun}, \citenamefont {Chen}, \citenamefont {Song}, \citenamefont {Xu},
		\citenamefont {Chu}, \citenamefont {Xiao} \emph {et~al.}}]{wang2016visible}%
	\BibitemOpen
	\bibfield  {author} {\bibinfo {author} {\bibfnamefont {B.}~\bibnamefont
			{Wang}}, \bibinfo {author} {\bibfnamefont {F.}~\bibnamefont {Dong}}, \bibinfo
		{author} {\bibfnamefont {Q.-T.}\ \bibnamefont {Li}}, \bibinfo {author}
		{\bibfnamefont {D.}~\bibnamefont {Yang}}, \bibinfo {author} {\bibfnamefont
			{C.}~\bibnamefont {Sun}}, \bibinfo {author} {\bibfnamefont {J.}~\bibnamefont
			{Chen}}, \bibinfo {author} {\bibfnamefont {Z.}~\bibnamefont {Song}}, \bibinfo
		{author} {\bibfnamefont {L.}~\bibnamefont {Xu}}, \bibinfo {author}
		{\bibfnamefont {W.}~\bibnamefont {Chu}}, \bibinfo {author} {\bibfnamefont
			{Y.-F.}\ \bibnamefont {Xiao}}, \emph {et~al.},\ }\bibfield  {title} {\bibinfo
		{title} {Visible-frequency dielectric metasurfaces for multiwavelength
			achromatic and highly dispersive holograms},\ }\href@noop {} {\bibfield
		{journal} {\bibinfo  {journal} {Nano Lett.}\ }\textbf {\bibinfo {volume}
			{16}},\ \bibinfo {pages} {5235} (\bibinfo {year} {2016})}\BibitemShut
	{NoStop}%
	\bibitem [{\citenamefont {Huang}\ \emph {et~al.}(2015)\citenamefont {Huang},
		\citenamefont {Chen}, \citenamefont {Tsai}, \citenamefont {Wu}, \citenamefont
		{Wang}, \citenamefont {Sun},\ and\ \citenamefont {Tsai}}]{huang2015aluminum}%
	\BibitemOpen
	\bibfield  {author} {\bibinfo {author} {\bibfnamefont {Y.-W.}\ \bibnamefont
			{Huang}}, \bibinfo {author} {\bibfnamefont {W.~T.}\ \bibnamefont {Chen}},
		\bibinfo {author} {\bibfnamefont {W.-Y.}\ \bibnamefont {Tsai}}, \bibinfo
		{author} {\bibfnamefont {P.~C.}\ \bibnamefont {Wu}}, \bibinfo {author}
		{\bibfnamefont {C.-M.}\ \bibnamefont {Wang}}, \bibinfo {author}
		{\bibfnamefont {G.}~\bibnamefont {Sun}},\ and\ \bibinfo {author}
		{\bibfnamefont {D.~P.}\ \bibnamefont {Tsai}},\ }\bibfield  {title} {\bibinfo
		{title} {Aluminum plasmonic multicolor meta-hologram},\ }\href@noop {}
	{\bibfield  {journal} {\bibinfo  {journal} {Nano Lett.}\ }\textbf {\bibinfo
			{volume} {15}},\ \bibinfo {pages} {3122} (\bibinfo {year}
		{2015})}\BibitemShut {NoStop}%
	\bibitem [{\citenamefont {Li}\ \emph {et~al.}(2021)\citenamefont {Li},
		\citenamefont {Hu}, \citenamefont {Shi}, \citenamefont {He}, \citenamefont
		{Li}, \citenamefont {Shang}, \citenamefont {Zhang}, \citenamefont {Fu},
		\citenamefont {Zhou}, \citenamefont {Xiong} \emph {et~al.}}]{li2021full}%
	\BibitemOpen
	\bibfield  {author} {\bibinfo {author} {\bibfnamefont {J.}~\bibnamefont
			{Li}}, \bibinfo {author} {\bibfnamefont {G.}~\bibnamefont {Hu}}, \bibinfo
		{author} {\bibfnamefont {L.}~\bibnamefont {Shi}}, \bibinfo {author}
		{\bibfnamefont {N.}~\bibnamefont {He}}, \bibinfo {author} {\bibfnamefont
			{D.}~\bibnamefont {Li}}, \bibinfo {author} {\bibfnamefont {Q.}~\bibnamefont
			{Shang}}, \bibinfo {author} {\bibfnamefont {Q.}~\bibnamefont {Zhang}},
		\bibinfo {author} {\bibfnamefont {H.}~\bibnamefont {Fu}}, \bibinfo {author}
		{\bibfnamefont {L.}~\bibnamefont {Zhou}}, \bibinfo {author} {\bibfnamefont
			{W.}~\bibnamefont {Xiong}}, \emph {et~al.},\ }\bibfield  {title} {\bibinfo
		{title} {Full-color enhanced second harmonic generation using rainbow
			trapping in ultrathin hyperbolic metamaterials},\ }\href@noop {} {\bibfield
		{journal} {\bibinfo  {journal} {Nat. Commun.}\ }\textbf {\bibinfo {volume}
			{12}},\ \bibinfo {pages} {6425} (\bibinfo {year} {2021})}\BibitemShut
	{NoStop}%
	\bibitem [{\citenamefont {Faraji-Dana}\ \emph {et~al.}(2018)\citenamefont
		{Faraji-Dana}, \citenamefont {Arbabi}, \citenamefont {Arbabi}, \citenamefont
		{Kamali}, \citenamefont {Kwon},\ and\ \citenamefont
		{Faraon}}]{faraji2018compact}%
	\BibitemOpen
	\bibfield  {author} {\bibinfo {author} {\bibfnamefont {M.}~\bibnamefont
			{Faraji-Dana}}, \bibinfo {author} {\bibfnamefont {E.}~\bibnamefont {Arbabi}},
		\bibinfo {author} {\bibfnamefont {A.}~\bibnamefont {Arbabi}}, \bibinfo
		{author} {\bibfnamefont {S.~M.}\ \bibnamefont {Kamali}}, \bibinfo {author}
		{\bibfnamefont {H.}~\bibnamefont {Kwon}},\ and\ \bibinfo {author}
		{\bibfnamefont {A.}~\bibnamefont {Faraon}},\ }\bibfield  {title} {\bibinfo
		{title} {Compact folded metasurface spectrometer},\ }\href@noop {} {\bibfield
		{journal} {\bibinfo  {journal} {Nat. Commun.}\ }\textbf {\bibinfo {volume}
			{9}},\ \bibinfo {pages} {4196} (\bibinfo {year} {2018})}\BibitemShut
	{NoStop}%
	\bibitem [{\citenamefont {Huang}\ \emph {et~al.}(2013)\citenamefont {Huang},
		\citenamefont {Chen}, \citenamefont {M{\"u}hlenbernd}, \citenamefont {Zhang},
		\citenamefont {Chen}, \citenamefont {Bai}, \citenamefont {Tan}, \citenamefont
		{Jin}, \citenamefont {Cheah}, \citenamefont {Qiu} \emph
		{et~al.}}]{huang2013three}%
	\BibitemOpen
	\bibfield  {author} {\bibinfo {author} {\bibfnamefont {L.}~\bibnamefont
			{Huang}}, \bibinfo {author} {\bibfnamefont {X.}~\bibnamefont {Chen}},
		\bibinfo {author} {\bibfnamefont {H.}~\bibnamefont {M{\"u}hlenbernd}},
		\bibinfo {author} {\bibfnamefont {H.}~\bibnamefont {Zhang}}, \bibinfo
		{author} {\bibfnamefont {S.}~\bibnamefont {Chen}}, \bibinfo {author}
		{\bibfnamefont {B.}~\bibnamefont {Bai}}, \bibinfo {author} {\bibfnamefont
			{Q.}~\bibnamefont {Tan}}, \bibinfo {author} {\bibfnamefont {G.}~\bibnamefont
			{Jin}}, \bibinfo {author} {\bibfnamefont {K.-W.}\ \bibnamefont {Cheah}},
		\bibinfo {author} {\bibfnamefont {C.-W.}\ \bibnamefont {Qiu}}, \emph
		{et~al.},\ }\bibfield  {title} {\bibinfo {title} {Three-dimensional optical
			holography using a plasmonic metasurface},\ }\href@noop {} {\bibfield
		{journal} {\bibinfo  {journal} {Nat. Commun.}\ }\textbf {\bibinfo {volume}
			{4}},\ \bibinfo {pages} {2808} (\bibinfo {year} {2013})}\BibitemShut
	{NoStop}%
	\bibitem [{\citenamefont {Zheng}\ \emph {et~al.}(2015)\citenamefont {Zheng},
		\citenamefont {M{\"u}hlenbernd}, \citenamefont {Kenney}, \citenamefont {Li},
		\citenamefont {Zentgraf},\ and\ \citenamefont
		{Zhang}}]{zheng2015metasurface}%
	\BibitemOpen
	\bibfield  {author} {\bibinfo {author} {\bibfnamefont {G.}~\bibnamefont
			{Zheng}}, \bibinfo {author} {\bibfnamefont {H.}~\bibnamefont
			{M{\"u}hlenbernd}}, \bibinfo {author} {\bibfnamefont {M.}~\bibnamefont
			{Kenney}}, \bibinfo {author} {\bibfnamefont {G.}~\bibnamefont {Li}}, \bibinfo
		{author} {\bibfnamefont {T.}~\bibnamefont {Zentgraf}},\ and\ \bibinfo
		{author} {\bibfnamefont {S.}~\bibnamefont {Zhang}},\ }\bibfield  {title}
	{\bibinfo {title} {Metasurface holograms reaching 80\% efficiency},\
	}\href@noop {} {\bibfield  {journal} {\bibinfo  {journal} {Nat.
				Nanotechnol.}\ }\textbf {\bibinfo {volume} {10}},\ \bibinfo {pages} {308}
		(\bibinfo {year} {2015})}\BibitemShut {NoStop}%
	\bibitem [{\citenamefont {Ye}\ \emph {et~al.}(2016)\citenamefont {Ye},
		\citenamefont {Zeuner}, \citenamefont {Li}, \citenamefont {Reineke},
		\citenamefont {He}, \citenamefont {Qiu}, \citenamefont {Liu}, \citenamefont
		{Wang}, \citenamefont {Zhang},\ and\ \citenamefont {Zentgraf}}]{ye2016spin}%
	\BibitemOpen
	\bibfield  {author} {\bibinfo {author} {\bibfnamefont {W.}~\bibnamefont
			{Ye}}, \bibinfo {author} {\bibfnamefont {F.}~\bibnamefont {Zeuner}}, \bibinfo
		{author} {\bibfnamefont {X.}~\bibnamefont {Li}}, \bibinfo {author}
		{\bibfnamefont {B.}~\bibnamefont {Reineke}}, \bibinfo {author} {\bibfnamefont
			{S.}~\bibnamefont {He}}, \bibinfo {author} {\bibfnamefont {C.-W.}\
			\bibnamefont {Qiu}}, \bibinfo {author} {\bibfnamefont {J.}~\bibnamefont
			{Liu}}, \bibinfo {author} {\bibfnamefont {Y.}~\bibnamefont {Wang}}, \bibinfo
		{author} {\bibfnamefont {S.}~\bibnamefont {Zhang}},\ and\ \bibinfo {author}
		{\bibfnamefont {T.}~\bibnamefont {Zentgraf}},\ }\bibfield  {title} {\bibinfo
		{title} {Spin and wavelength multiplexed nonlinear metasurface holography},\
	}\href@noop {} {\bibfield  {journal} {\bibinfo  {journal} {Nat. Commun.}\
		}\textbf {\bibinfo {volume} {7}},\ \bibinfo {pages} {11930} (\bibinfo {year}
		{2016})}\BibitemShut {NoStop}%
	\bibitem [{\citenamefont {Wan}\ \emph {et~al.}(2017)\citenamefont {Wan},
		\citenamefont {Gao},\ and\ \citenamefont {Yang}}]{wan2017metasurface}%
	\BibitemOpen
	\bibfield  {author} {\bibinfo {author} {\bibfnamefont {W.}~\bibnamefont
			{Wan}}, \bibinfo {author} {\bibfnamefont {J.}~\bibnamefont {Gao}},\ and\
		\bibinfo {author} {\bibfnamefont {X.}~\bibnamefont {Yang}},\ }\bibfield
	{title} {\bibinfo {title} {Metasurface holograms for holographic imaging},\
	}\href@noop {} {\bibfield  {journal} {\bibinfo  {journal} {Adv. Opt. Mater.}\
		}\textbf {\bibinfo {volume} {5}},\ \bibinfo {pages} {1700541} (\bibinfo
		{year} {2017})}\BibitemShut {NoStop}%
	\bibitem [{\citenamefont {Yue}\ \emph {et~al.}(2018)\citenamefont {Yue},
		\citenamefont {Zhang}, \citenamefont {Zang}, \citenamefont {Wen},
		\citenamefont {Gerardot}, \citenamefont {Zhang},\ and\ \citenamefont
		{Chen}}]{yue2018high}%
	\BibitemOpen
	\bibfield  {author} {\bibinfo {author} {\bibfnamefont {F.}~\bibnamefont
			{Yue}}, \bibinfo {author} {\bibfnamefont {C.}~\bibnamefont {Zhang}}, \bibinfo
		{author} {\bibfnamefont {X.-F.}\ \bibnamefont {Zang}}, \bibinfo {author}
		{\bibfnamefont {D.}~\bibnamefont {Wen}}, \bibinfo {author} {\bibfnamefont
			{B.~D.}\ \bibnamefont {Gerardot}}, \bibinfo {author} {\bibfnamefont
			{S.}~\bibnamefont {Zhang}},\ and\ \bibinfo {author} {\bibfnamefont
			{X.}~\bibnamefont {Chen}},\ }\bibfield  {title} {\bibinfo {title}
		{High-resolution grayscale image hidden in a laser beam},\ }\href@noop {}
	{\bibfield  {journal} {\bibinfo  {journal} {Light Sci. Appl.}\ }\textbf
		{\bibinfo {volume} {7}},\ \bibinfo {pages} {17129} (\bibinfo {year}
		{2018})}\BibitemShut {NoStop}%
	\bibitem [{\citenamefont {Dai}\ \emph {et~al.}(2020{\natexlab{a}})\citenamefont
		{Dai}, \citenamefont {Zhou}, \citenamefont {Deng}, \citenamefont {Deng},
		\citenamefont {Li},\ and\ \citenamefont {Zheng}}]{dai2020dual}%
	\BibitemOpen
	\bibfield  {author} {\bibinfo {author} {\bibfnamefont {Q.}~\bibnamefont
			{Dai}}, \bibinfo {author} {\bibfnamefont {N.}~\bibnamefont {Zhou}}, \bibinfo
		{author} {\bibfnamefont {L.}~\bibnamefont {Deng}}, \bibinfo {author}
		{\bibfnamefont {J.}~\bibnamefont {Deng}}, \bibinfo {author} {\bibfnamefont
			{Z.}~\bibnamefont {Li}},\ and\ \bibinfo {author} {\bibfnamefont
			{G.}~\bibnamefont {Zheng}},\ }\bibfield  {title} {\bibinfo {title}
		{Dual-channel binary gray-image display enabled with malus-assisted
			metasurfaces},\ }\href@noop {} {\bibfield  {journal} {\bibinfo  {journal}
			{Phys. Rev. Applied}\ }\textbf {\bibinfo {volume} {14}},\ \bibinfo {pages}
		{034002} (\bibinfo {year} {2020}{\natexlab{a}})}\BibitemShut {NoStop}%
	\bibitem [{\citenamefont {Deng}\ \emph
		{et~al.}(2020{\natexlab{b}})\citenamefont {Deng}, \citenamefont {Deng},
		\citenamefont {Guan}, \citenamefont {Tao}, \citenamefont {Li}, \citenamefont
		{Li}, \citenamefont {Li}, \citenamefont {Yu},\ and\ \citenamefont
		{Zheng}}]{deng2020multiplexed}%
	\BibitemOpen
	\bibfield  {author} {\bibinfo {author} {\bibfnamefont {J.}~\bibnamefont
			{Deng}}, \bibinfo {author} {\bibfnamefont {L.}~\bibnamefont {Deng}}, \bibinfo
		{author} {\bibfnamefont {Z.}~\bibnamefont {Guan}}, \bibinfo {author}
		{\bibfnamefont {J.}~\bibnamefont {Tao}}, \bibinfo {author} {\bibfnamefont
			{G.}~\bibnamefont {Li}}, \bibinfo {author} {\bibfnamefont {Z.}~\bibnamefont
			{Li}}, \bibinfo {author} {\bibfnamefont {Z.}~\bibnamefont {Li}}, \bibinfo
		{author} {\bibfnamefont {S.}~\bibnamefont {Yu}},\ and\ \bibinfo {author}
		{\bibfnamefont {G.}~\bibnamefont {Zheng}},\ }\bibfield  {title} {\bibinfo
		{title} {Multiplexed anticounterfeiting meta-image displays with single-sized
			nanostructures},\ }\href@noop {} {\bibfield  {journal} {\bibinfo  {journal}
			{Nano Lett.}\ }\textbf {\bibinfo {volume} {20}},\ \bibinfo {pages} {1830}
		(\bibinfo {year} {2020}{\natexlab{b}})}\BibitemShut {NoStop}%
	\bibitem [{\citenamefont {Goh}\ \emph {et~al.}(2014)\citenamefont {Goh},
		\citenamefont {Zheng}, \citenamefont {Tan}, \citenamefont {Zhang},
		\citenamefont {Kumar}, \citenamefont {Qiu},\ and\ \citenamefont
		{Yang}}]{goh2014three}%
	\BibitemOpen
	\bibfield  {author} {\bibinfo {author} {\bibfnamefont {X.~M.}\ \bibnamefont
			{Goh}}, \bibinfo {author} {\bibfnamefont {Y.}~\bibnamefont {Zheng}}, \bibinfo
		{author} {\bibfnamefont {S.~J.}\ \bibnamefont {Tan}}, \bibinfo {author}
		{\bibfnamefont {L.}~\bibnamefont {Zhang}}, \bibinfo {author} {\bibfnamefont
			{K.}~\bibnamefont {Kumar}}, \bibinfo {author} {\bibfnamefont {C.-W.}\
			\bibnamefont {Qiu}},\ and\ \bibinfo {author} {\bibfnamefont {J.~K.}\
			\bibnamefont {Yang}},\ }\bibfield  {title} {\bibinfo {title}
		{Three-dimensional plasmonic stereoscopic prints in full colour},\
	}\href@noop {} {\bibfield  {journal} {\bibinfo  {journal} {Nat. Commun.}\
		}\textbf {\bibinfo {volume} {5}},\ \bibinfo {pages} {5361} (\bibinfo {year}
		{2014})}\BibitemShut {NoStop}%
	\bibitem [{\citenamefont {Dalloz}\ \emph {et~al.}(2022)\citenamefont {Dalloz},
		\citenamefont {Le}, \citenamefont {Hebert}, \citenamefont {Eles},
		\citenamefont {Flores~Figueroa}, \citenamefont {Hubert}, \citenamefont {Ma},
		\citenamefont {Sharma}, \citenamefont {Vocanson}, \citenamefont {Ayala} \emph
		{et~al.}}]{dalloz2022anti}%
	\BibitemOpen
	\bibfield  {author} {\bibinfo {author} {\bibfnamefont {N.}~\bibnamefont
			{Dalloz}}, \bibinfo {author} {\bibfnamefont {V.~D.}\ \bibnamefont {Le}},
		\bibinfo {author} {\bibfnamefont {M.}~\bibnamefont {Hebert}}, \bibinfo
		{author} {\bibfnamefont {B.}~\bibnamefont {Eles}}, \bibinfo {author}
		{\bibfnamefont {M.~A.}\ \bibnamefont {Flores~Figueroa}}, \bibinfo {author}
		{\bibfnamefont {C.}~\bibnamefont {Hubert}}, \bibinfo {author} {\bibfnamefont
			{H.}~\bibnamefont {Ma}}, \bibinfo {author} {\bibfnamefont {N.}~\bibnamefont
			{Sharma}}, \bibinfo {author} {\bibfnamefont {F.}~\bibnamefont {Vocanson}},
		\bibinfo {author} {\bibfnamefont {S.}~\bibnamefont {Ayala}}, \emph {et~al.},\
	}\bibfield  {title} {\bibinfo {title} {Anti-counterfeiting white light
			printed image multiplexing by fast nanosecond laser processing},\ }\href@noop
	{} {\bibfield  {journal} {\bibinfo  {journal} {Adv. Mater.}\ }\textbf
		{\bibinfo {volume} {34}},\ \bibinfo {pages} {2104054} (\bibinfo {year}
		{2022})}\BibitemShut {NoStop}%
	\bibitem [{\citenamefont {Wang}\ \emph {et~al.}(2020)\citenamefont {Wang},
		\citenamefont {Niu}, \citenamefont {Liang}, \citenamefont {Li}, \citenamefont
		{Hua}, \citenamefont {Shi},\ and\ \citenamefont {Xie}}]{wang2020complete}%
	\BibitemOpen
	\bibfield  {author} {\bibinfo {author} {\bibfnamefont {E.}~\bibnamefont
			{Wang}}, \bibinfo {author} {\bibfnamefont {J.}~\bibnamefont {Niu}}, \bibinfo
		{author} {\bibfnamefont {Y.}~\bibnamefont {Liang}}, \bibinfo {author}
		{\bibfnamefont {H.}~\bibnamefont {Li}}, \bibinfo {author} {\bibfnamefont
			{Y.}~\bibnamefont {Hua}}, \bibinfo {author} {\bibfnamefont {L.}~\bibnamefont
			{Shi}},\ and\ \bibinfo {author} {\bibfnamefont {C.}~\bibnamefont {Xie}},\
	}\bibfield  {title} {\bibinfo {title} {Complete control of multichannel,
			angle-multiplexed, and arbitrary spatially varying polarization fields},\
	}\href@noop {} {\bibfield  {journal} {\bibinfo  {journal} {Adv. Opt. Mater.}\
		}\textbf {\bibinfo {volume} {8}},\ \bibinfo {pages} {1901674} (\bibinfo
		{year} {2020})}\BibitemShut {NoStop}%
	\bibitem [{\citenamefont {Arbabi}\ \emph
		{et~al.}(2015{\natexlab{a}})\citenamefont {Arbabi}, \citenamefont {Horie},
		\citenamefont {Ball}, \citenamefont {Bagheri},\ and\ \citenamefont
		{Faraon}}]{arbabi2015subwavelength}%
	\BibitemOpen
	\bibfield  {author} {\bibinfo {author} {\bibfnamefont {A.}~\bibnamefont
			{Arbabi}}, \bibinfo {author} {\bibfnamefont {Y.}~\bibnamefont {Horie}},
		\bibinfo {author} {\bibfnamefont {A.~J.}\ \bibnamefont {Ball}}, \bibinfo
		{author} {\bibfnamefont {M.}~\bibnamefont {Bagheri}},\ and\ \bibinfo {author}
		{\bibfnamefont {A.}~\bibnamefont {Faraon}},\ }\bibfield  {title} {\bibinfo
		{title} {Subwavelength-thick lenses with high numerical apertures and large
			efficiency based on high-contrast transmitarrays},\ }\href@noop {} {\bibfield
		{journal} {\bibinfo  {journal} {Nat. Commun.}\ }\textbf {\bibinfo {volume}
			{6}},\ \bibinfo {pages} {7069} (\bibinfo {year}
		{2015}{\natexlab{a}})}\BibitemShut {NoStop}%
	\bibitem [{\citenamefont {Arbabi}\ \emph {et~al.}(2016)\citenamefont {Arbabi},
		\citenamefont {Arbabi}, \citenamefont {Kamali}, \citenamefont {Horie},
		\citenamefont {Han},\ and\ \citenamefont {Faraon}}]{arbabi2016miniature}%
	\BibitemOpen
	\bibfield  {author} {\bibinfo {author} {\bibfnamefont {A.}~\bibnamefont
			{Arbabi}}, \bibinfo {author} {\bibfnamefont {E.}~\bibnamefont {Arbabi}},
		\bibinfo {author} {\bibfnamefont {S.~M.}\ \bibnamefont {Kamali}}, \bibinfo
		{author} {\bibfnamefont {Y.}~\bibnamefont {Horie}}, \bibinfo {author}
		{\bibfnamefont {S.}~\bibnamefont {Han}},\ and\ \bibinfo {author}
		{\bibfnamefont {A.}~\bibnamefont {Faraon}},\ }\bibfield  {title} {\bibinfo
		{title} {Miniature optical planar camera based on a wide-angle metasurface
			doublet corrected for monochromatic aberrations},\ }\href@noop {} {\bibfield
		{journal} {\bibinfo  {journal} {Nat. Commun.}\ }\textbf {\bibinfo {volume}
			{7}},\ \bibinfo {pages} {13682} (\bibinfo {year} {2016})}\BibitemShut
	{NoStop}%
	\bibitem [{\citenamefont {Chen}\ \emph {et~al.}(2017)\citenamefont {Chen},
		\citenamefont {Feng}, \citenamefont {Monticone}, \citenamefont {Zhao},
		\citenamefont {Zhu}, \citenamefont {Jiang}, \citenamefont {Zhang},
		\citenamefont {Kim}, \citenamefont {Ding}, \citenamefont {Zhang} \emph
		{et~al.}}]{chen2017reconfigurable}%
	\BibitemOpen
	\bibfield  {author} {\bibinfo {author} {\bibfnamefont {K.}~\bibnamefont
			{Chen}}, \bibinfo {author} {\bibfnamefont {Y.}~\bibnamefont {Feng}}, \bibinfo
		{author} {\bibfnamefont {F.}~\bibnamefont {Monticone}}, \bibinfo {author}
		{\bibfnamefont {J.}~\bibnamefont {Zhao}}, \bibinfo {author} {\bibfnamefont
			{B.}~\bibnamefont {Zhu}}, \bibinfo {author} {\bibfnamefont {T.}~\bibnamefont
			{Jiang}}, \bibinfo {author} {\bibfnamefont {L.}~\bibnamefont {Zhang}},
		\bibinfo {author} {\bibfnamefont {Y.}~\bibnamefont {Kim}}, \bibinfo {author}
		{\bibfnamefont {X.}~\bibnamefont {Ding}}, \bibinfo {author} {\bibfnamefont
			{S.}~\bibnamefont {Zhang}}, \emph {et~al.},\ }\bibfield  {title} {\bibinfo
		{title} {A reconfigurable active huygens' metalens},\ }\href@noop {}
	{\bibfield  {journal} {\bibinfo  {journal} {Adv. Mater.}\ }\textbf {\bibinfo
			{volume} {29}},\ \bibinfo {pages} {1606422} (\bibinfo {year}
		{2017})}\BibitemShut {NoStop}%
	\bibitem [{\citenamefont {Chen}\ \emph {et~al.}(2018)\citenamefont {Chen},
		\citenamefont {Zhu}, \citenamefont {Sanjeev}, \citenamefont {Khorasaninejad},
		\citenamefont {Shi}, \citenamefont {Lee},\ and\ \citenamefont
		{Capasso}}]{chen2018broadband}%
	\BibitemOpen
	\bibfield  {author} {\bibinfo {author} {\bibfnamefont {W.~T.}\ \bibnamefont
			{Chen}}, \bibinfo {author} {\bibfnamefont {A.~Y.}\ \bibnamefont {Zhu}},
		\bibinfo {author} {\bibfnamefont {V.}~\bibnamefont {Sanjeev}}, \bibinfo
		{author} {\bibfnamefont {M.}~\bibnamefont {Khorasaninejad}}, \bibinfo
		{author} {\bibfnamefont {Z.}~\bibnamefont {Shi}}, \bibinfo {author}
		{\bibfnamefont {E.}~\bibnamefont {Lee}},\ and\ \bibinfo {author}
		{\bibfnamefont {F.}~\bibnamefont {Capasso}},\ }\bibfield  {title} {\bibinfo
		{title} {A broadband achromatic metalens for focusing and imaging in the
			visible},\ }\href@noop {} {\bibfield  {journal} {\bibinfo  {journal} {Nat.
				Nanotechnol.}\ }\textbf {\bibinfo {volume} {13}},\ \bibinfo {pages} {220}
		(\bibinfo {year} {2018})}\BibitemShut {NoStop}%
	\bibitem [{\citenamefont {Wang}\ \emph {et~al.}(2021)\citenamefont {Wang},
		\citenamefont {Fan}, \citenamefont {Xu} \emph {et~al.}}]{wang2021design}%
	\BibitemOpen
	\bibfield  {author} {\bibinfo {author} {\bibfnamefont {Y.}~\bibnamefont
			{Wang}}, \bibinfo {author} {\bibfnamefont {Q.}~\bibnamefont {Fan}}, \bibinfo
		{author} {\bibfnamefont {T.}~\bibnamefont {Xu}}, \emph {et~al.},\ }\bibfield
	{title} {\bibinfo {title} {Design of high efficiency achromatic metalens with
			large operation bandwidth using bilayer architecture},\ }\href@noop {}
	{\bibfield  {journal} {\bibinfo  {journal} {Opto-Electron. Adv.}\ }\textbf
		{\bibinfo {volume} {4}},\ \bibinfo {pages} {200008} (\bibinfo {year}
		{2021})}\BibitemShut {NoStop}%
	\bibitem [{\citenamefont {Tseng}\ \emph {et~al.}(2021)\citenamefont {Tseng},
		\citenamefont {Colburn}, \citenamefont {Whitehead}, \citenamefont {Huang},
		\citenamefont {Baek}, \citenamefont {Majumdar},\ and\ \citenamefont
		{Heide}}]{tseng2021neural}%
	\BibitemOpen
	\bibfield  {author} {\bibinfo {author} {\bibfnamefont {E.}~\bibnamefont
			{Tseng}}, \bibinfo {author} {\bibfnamefont {S.}~\bibnamefont {Colburn}},
		\bibinfo {author} {\bibfnamefont {J.}~\bibnamefont {Whitehead}}, \bibinfo
		{author} {\bibfnamefont {L.}~\bibnamefont {Huang}}, \bibinfo {author}
		{\bibfnamefont {S.-H.}\ \bibnamefont {Baek}}, \bibinfo {author}
		{\bibfnamefont {A.}~\bibnamefont {Majumdar}},\ and\ \bibinfo {author}
		{\bibfnamefont {F.}~\bibnamefont {Heide}},\ }\bibfield  {title} {\bibinfo
		{title} {Neural nano-optics for high-quality thin lens imaging},\ }\href@noop
	{} {\bibfield  {journal} {\bibinfo  {journal} {Nat. Commun.}\ }\textbf
		{\bibinfo {volume} {12}},\ \bibinfo {pages} {6493} (\bibinfo {year}
		{2021})}\BibitemShut {NoStop}%
	\bibitem [{\citenamefont {Zhang}\ \emph {et~al.}(2023)\citenamefont {Zhang},
		\citenamefont {Fang}, \citenamefont {Zhao}, \citenamefont {Li}, \citenamefont
		{Shen}, \citenamefont {Hong},\ and\ \citenamefont
		{Jing}}]{zhang2023terahertz}%
	\BibitemOpen
	\bibfield  {author} {\bibinfo {author} {\bibfnamefont {P.}~\bibnamefont
			{Zhang}}, \bibinfo {author} {\bibfnamefont {B.}~\bibnamefont {Fang}},
		\bibinfo {author} {\bibfnamefont {T.}~\bibnamefont {Zhao}}, \bibinfo {author}
		{\bibfnamefont {C.}~\bibnamefont {Li}}, \bibinfo {author} {\bibfnamefont
			{C.}~\bibnamefont {Shen}}, \bibinfo {author} {\bibfnamefont {Z.}~\bibnamefont
			{Hong}},\ and\ \bibinfo {author} {\bibfnamefont {X.}~\bibnamefont {Jing}},\
	}\bibfield  {title} {\bibinfo {title} {Terahertz wave tunable metalens based
			on phase change material coded metasurface},\ }\href@noop {} {\bibfield
		{journal} {\bibinfo  {journal} {J. Lightwave Technol.}\ }\textbf {\bibinfo
			{volume} {41}},\ \bibinfo {pages} {7162} (\bibinfo {year}
		{2023})}\BibitemShut {NoStop}%
	\bibitem [{\citenamefont {Liu}\ \emph {et~al.}(2024)\citenamefont {Liu},
		\citenamefont {Qiu}, \citenamefont {Xu}, \citenamefont {Qin}, \citenamefont
		{Wan}, \citenamefont {Yu}, \citenamefont {Liu}, \citenamefont {Huang},\ and\
		\citenamefont {Xiao}}]{liu2024edge}%
	\BibitemOpen
	\bibfield  {author} {\bibinfo {author} {\bibfnamefont {T.}~\bibnamefont
			{Liu}}, \bibinfo {author} {\bibfnamefont {J.}~\bibnamefont {Qiu}}, \bibinfo
		{author} {\bibfnamefont {L.}~\bibnamefont {Xu}}, \bibinfo {author}
		{\bibfnamefont {M.}~\bibnamefont {Qin}}, \bibinfo {author} {\bibfnamefont
			{L.}~\bibnamefont {Wan}}, \bibinfo {author} {\bibfnamefont {T.}~\bibnamefont
			{Yu}}, \bibinfo {author} {\bibfnamefont {Q.}~\bibnamefont {Liu}}, \bibinfo
		{author} {\bibfnamefont {L.}~\bibnamefont {Huang}},\ and\ \bibinfo {author}
		{\bibfnamefont {S.}~\bibnamefont {Xiao}},\ }\bibfield  {title} {\bibinfo
		{title} {Edge detection imaging by quasi-bound states in the continuum},\
	}\href@noop {} {\bibfield  {journal} {\bibinfo  {journal} {Nano Lett.}\
		}\textbf {\bibinfo {volume} {24}},\ \bibinfo {pages} {14466} (\bibinfo {year}
		{2024})}\BibitemShut {NoStop}%
	\bibitem [{\citenamefont {Liu}\ \emph {et~al.}(2025)\citenamefont {Liu},
		\citenamefont {Qiu}, \citenamefont {Yu}, \citenamefont {Liu}, \citenamefont
		{Li},\ and\ \citenamefont {Xiao}}]{Liu2025}%
	\BibitemOpen
	\bibfield  {author} {\bibinfo {author} {\bibfnamefont {T.}~\bibnamefont
			{Liu}}, \bibinfo {author} {\bibfnamefont {J.}~\bibnamefont {Qiu}}, \bibinfo
		{author} {\bibfnamefont {T.}~\bibnamefont {Yu}}, \bibinfo {author}
		{\bibfnamefont {Q.}~\bibnamefont {Liu}}, \bibinfo {author} {\bibfnamefont
			{J.}~\bibnamefont {Li}},\ and\ \bibinfo {author} {\bibfnamefont
			{S.}~\bibnamefont {Xiao}},\ }\bibfield  {title} {\bibinfo {title}
		{Phase-change metasurfaces for reconfigurable image processing},\ }\href
	{https://doi.org/10.1063/5.0248307} {\bibfield  {journal} {\bibinfo
			{journal} {Appl. Phys. Lett.}\ }\textbf {\bibinfo {volume} {126}},\ \bibinfo
		{pages} {081702} (\bibinfo {year} {2025})}\BibitemShut {NoStop}%
	\bibitem [{\citenamefont {Zhang}\ \emph {et~al.}(2025)\citenamefont {Zhang},
		\citenamefont {Wang}, \citenamefont {Qiu}, \citenamefont {Yang},
		\citenamefont {Liu}, \citenamefont {Xiao}, \citenamefont {Staude},
		\citenamefont {Pertsch}, \citenamefont {Wang},\ and\ \citenamefont
		{Zou}}]{Zhang2025}%
	\BibitemOpen
	\bibfield  {author} {\bibinfo {author} {\bibfnamefont {K.}~\bibnamefont
			{Zhang}}, \bibinfo {author} {\bibfnamefont {S.}~\bibnamefont {Wang}},
		\bibinfo {author} {\bibfnamefont {J.}~\bibnamefont {Qiu}}, \bibinfo {author}
		{\bibfnamefont {M.}~\bibnamefont {Yang}}, \bibinfo {author} {\bibfnamefont
			{T.}~\bibnamefont {Liu}}, \bibinfo {author} {\bibfnamefont {S.}~\bibnamefont
			{Xiao}}, \bibinfo {author} {\bibfnamefont {I.}~\bibnamefont {Staude}},
		\bibinfo {author} {\bibfnamefont {T.}~\bibnamefont {Pertsch}}, \bibinfo
		{author} {\bibfnamefont {Y.}~\bibnamefont {Wang}},\ and\ \bibinfo {author}
		{\bibfnamefont {C.}~\bibnamefont {Zou}},\ }\bibfield  {title} {\bibinfo
		{title} {Momentum\space tunable metasurfaces for switchable image
			processing},\ }\href {https://doi.org/10.1002/adom.202500352} {\bibfield
		{journal} {\bibinfo  {journal} {Adv. Opt. Mater.}\ }\textbf {\bibinfo
			{volume} {13}},\ \bibinfo {pages} {2500352} (\bibinfo {year}
		{2025})}\BibitemShut {NoStop}%
	\bibitem [{\citenamefont {Zong}\ \emph {et~al.}(2025)\citenamefont {Zong},
		\citenamefont {Zhang}, \citenamefont {Liu}, \citenamefont {Fan},
		\citenamefont {Lv}, \citenamefont {Wang},\ and\ \citenamefont
		{Xu}}]{Zong2025}%
	\BibitemOpen
	\bibfield  {author} {\bibinfo {author} {\bibfnamefont {M.}~\bibnamefont
			{Zong}}, \bibinfo {author} {\bibfnamefont {S.}~\bibnamefont {Zhang}},
		\bibinfo {author} {\bibfnamefont {Y.}~\bibnamefont {Liu}}, \bibinfo {author}
		{\bibfnamefont {Y.}~\bibnamefont {Fan}}, \bibinfo {author} {\bibfnamefont
			{J.}~\bibnamefont {Lv}}, \bibinfo {author} {\bibfnamefont {X.}~\bibnamefont
			{Wang}},\ and\ \bibinfo {author} {\bibfnamefont {Z.}~\bibnamefont {Xu}},\
	}\bibfield  {title} {\bibinfo {title} {Polarization- and
			wavelength-multiplexed multichannel optical differentiators for edge
			detection},\ }\href {https://doi.org/10.1021/acsphotonics.5c00742} {\bibfield
		{journal} {\bibinfo  {journal} {ACS Photonics}\ }\textbf {\bibinfo {volume}
			{12}},\ \bibinfo {pages} {4366} (\bibinfo {year} {2025})}\BibitemShut
	{NoStop}%
	\bibitem [{\citenamefont {Shalaev}\ \emph {et~al.}(2015)\citenamefont
		{Shalaev}, \citenamefont {Sun}, \citenamefont {Tsukernik}, \citenamefont
		{Pandey}, \citenamefont {Nikolskiy},\ and\ \citenamefont
		{Litchinitser}}]{shalaev2015high}%
	\BibitemOpen
	\bibfield  {author} {\bibinfo {author} {\bibfnamefont {M.~I.}\ \bibnamefont
			{Shalaev}}, \bibinfo {author} {\bibfnamefont {J.}~\bibnamefont {Sun}},
		\bibinfo {author} {\bibfnamefont {A.}~\bibnamefont {Tsukernik}}, \bibinfo
		{author} {\bibfnamefont {A.}~\bibnamefont {Pandey}}, \bibinfo {author}
		{\bibfnamefont {K.}~\bibnamefont {Nikolskiy}},\ and\ \bibinfo {author}
		{\bibfnamefont {N.~M.}\ \bibnamefont {Litchinitser}},\ }\bibfield  {title}
	{\bibinfo {title} {High-efficiency all-dielectric metasurfaces for
			ultracompact beam manipulation in transmission mode},\ }\href@noop {}
	{\bibfield  {journal} {\bibinfo  {journal} {Nano Lett.}\ }\textbf {\bibinfo
			{volume} {15}},\ \bibinfo {pages} {6261} (\bibinfo {year}
		{2015})}\BibitemShut {NoStop}%
	\bibitem [{\citenamefont {Mehmood}\ \emph {et~al.}(2016)\citenamefont
		{Mehmood}, \citenamefont {Mei}, \citenamefont {Hussain}, \citenamefont
		{Huang}, \citenamefont {Siew}, \citenamefont {Zhang}, \citenamefont {Zhang},
		\citenamefont {Ling}, \citenamefont {Liu}, \citenamefont {Teng} \emph
		{et~al.}}]{mehmood2016visible}%
	\BibitemOpen
	\bibfield  {author} {\bibinfo {author} {\bibfnamefont {M.}~\bibnamefont
			{Mehmood}}, \bibinfo {author} {\bibfnamefont {S.}~\bibnamefont {Mei}},
		\bibinfo {author} {\bibfnamefont {S.}~\bibnamefont {Hussain}}, \bibinfo
		{author} {\bibfnamefont {K.}~\bibnamefont {Huang}}, \bibinfo {author}
		{\bibfnamefont {S.}~\bibnamefont {Siew}}, \bibinfo {author} {\bibfnamefont
			{L.}~\bibnamefont {Zhang}}, \bibinfo {author} {\bibfnamefont
			{T.}~\bibnamefont {Zhang}}, \bibinfo {author} {\bibfnamefont
			{X.}~\bibnamefont {Ling}}, \bibinfo {author} {\bibfnamefont {H.}~\bibnamefont
			{Liu}}, \bibinfo {author} {\bibfnamefont {J.}~\bibnamefont {Teng}}, \emph
		{et~al.},\ }\bibfield  {title} {\bibinfo {title} {Visible-frequency
			metasurface for structuring and spatially multiplexing optical vortices},\
	}\href@noop {} {\bibfield  {journal} {\bibinfo  {journal} {Adv. Mater}\
		}\textbf {\bibinfo {volume} {28}},\ \bibinfo {pages} {2533} (\bibinfo {year}
		{2016})}\BibitemShut {NoStop}%
	\bibitem [{\citenamefont {Ren}\ \emph {et~al.}(2019)\citenamefont {Ren},
		\citenamefont {Briere}, \citenamefont {Fang}, \citenamefont {Ni},
		\citenamefont {Sawant}, \citenamefont {H{\'e}ron}, \citenamefont {Chenot},
		\citenamefont {V{\'e}zian}, \citenamefont {Damilano}, \citenamefont
		{Br{\"a}ndli} \emph {et~al.}}]{ren2019metasurface}%
	\BibitemOpen
	\bibfield  {author} {\bibinfo {author} {\bibfnamefont {H.}~\bibnamefont
			{Ren}}, \bibinfo {author} {\bibfnamefont {G.}~\bibnamefont {Briere}},
		\bibinfo {author} {\bibfnamefont {X.}~\bibnamefont {Fang}}, \bibinfo {author}
		{\bibfnamefont {P.}~\bibnamefont {Ni}}, \bibinfo {author} {\bibfnamefont
			{R.}~\bibnamefont {Sawant}}, \bibinfo {author} {\bibfnamefont
			{S.}~\bibnamefont {H{\'e}ron}}, \bibinfo {author} {\bibfnamefont
			{S.}~\bibnamefont {Chenot}}, \bibinfo {author} {\bibfnamefont
			{S.}~\bibnamefont {V{\'e}zian}}, \bibinfo {author} {\bibfnamefont
			{B.}~\bibnamefont {Damilano}}, \bibinfo {author} {\bibfnamefont
			{V.}~\bibnamefont {Br{\"a}ndli}}, \emph {et~al.},\ }\bibfield  {title}
	{\bibinfo {title} {Metasurface orbital angular momentum holography},\
	}\href@noop {} {\bibfield  {journal} {\bibinfo  {journal} {Nat. Commun.}\
		}\textbf {\bibinfo {volume} {10}},\ \bibinfo {pages} {2986} (\bibinfo {year}
		{2019})}\BibitemShut {NoStop}%
	\bibitem [{\citenamefont {Bao}\ \emph {et~al.}(2020)\citenamefont {Bao},
		\citenamefont {Ni},\ and\ \citenamefont {Qiu}}]{bao2020minimalist}%
	\BibitemOpen
	\bibfield  {author} {\bibinfo {author} {\bibfnamefont {Y.}~\bibnamefont
			{Bao}}, \bibinfo {author} {\bibfnamefont {J.}~\bibnamefont {Ni}},\ and\
		\bibinfo {author} {\bibfnamefont {C.-W.}\ \bibnamefont {Qiu}},\ }\bibfield
	{title} {\bibinfo {title} {A minimalist single-layer metasurface for
			arbitrary and full control of vector vortex beams},\ }\href@noop {}
	{\bibfield  {journal} {\bibinfo  {journal} {Adv. Mater.}\ }\textbf {\bibinfo
			{volume} {32}},\ \bibinfo {pages} {1905659} (\bibinfo {year}
		{2020})}\BibitemShut {NoStop}%
	\bibitem [{\citenamefont {Shi}\ \emph {et~al.}(2023)\citenamefont {Shi},
		\citenamefont {Wang}, \citenamefont {Li}, \citenamefont {Yi}, \citenamefont
		{Liu}, \citenamefont {Zhang},\ and\ \citenamefont {Xu}}]{shi2023guided}%
	\BibitemOpen
	\bibfield  {author} {\bibinfo {author} {\bibfnamefont {H.}~\bibnamefont
			{Shi}}, \bibinfo {author} {\bibfnamefont {L.}~\bibnamefont {Wang}}, \bibinfo
		{author} {\bibfnamefont {G.}~\bibnamefont {Li}}, \bibinfo {author}
		{\bibfnamefont {J.}~\bibnamefont {Yi}}, \bibinfo {author} {\bibfnamefont
			{H.}~\bibnamefont {Liu}}, \bibinfo {author} {\bibfnamefont {A.}~\bibnamefont
			{Zhang}},\ and\ \bibinfo {author} {\bibfnamefont {Z.}~\bibnamefont {Xu}},\
	}\bibfield  {title} {\bibinfo {title} {Guided-wave inspired metasurfaces for
			multifunctional vortex beam generation and manipulation},\ }\href@noop {}
	{\bibfield  {journal} {\bibinfo  {journal} {J. Lightwave Technol.}\ }\textbf
		{\bibinfo {volume} {41}},\ \bibinfo {pages} {2094} (\bibinfo {year}
		{2023})}\BibitemShut {NoStop}%
	\bibitem [{\citenamefont {Arbabi}\ \emph
		{et~al.}(2015{\natexlab{b}})\citenamefont {Arbabi}, \citenamefont {Horie},
		\citenamefont {Bagheri},\ and\ \citenamefont
		{Faraon}}]{arbabi2015dielectric}%
	\BibitemOpen
	\bibfield  {author} {\bibinfo {author} {\bibfnamefont {A.}~\bibnamefont
			{Arbabi}}, \bibinfo {author} {\bibfnamefont {Y.}~\bibnamefont {Horie}},
		\bibinfo {author} {\bibfnamefont {M.}~\bibnamefont {Bagheri}},\ and\ \bibinfo
		{author} {\bibfnamefont {A.}~\bibnamefont {Faraon}},\ }\bibfield  {title}
	{\bibinfo {title} {Dielectric metasurfaces for complete control of phase and
			polarization with subwavelength spatial resolution and high transmission},\
	}\href@noop {} {\bibfield  {journal} {\bibinfo  {journal} {Nat.
				Nanotechnol.}\ }\textbf {\bibinfo {volume} {10}},\ \bibinfo {pages} {937}
		(\bibinfo {year} {2015}{\natexlab{b}})}\BibitemShut {NoStop}%
	\bibitem [{\citenamefont {Overvig}\ \emph {et~al.}(2019)\citenamefont
		{Overvig}, \citenamefont {Shrestha}, \citenamefont {Malek}, \citenamefont
		{Lu}, \citenamefont {Stein}, \citenamefont {Zheng},\ and\ \citenamefont
		{Yu}}]{overvig2019dielectric}%
	\BibitemOpen
	\bibfield  {author} {\bibinfo {author} {\bibfnamefont {A.~C.}\ \bibnamefont
			{Overvig}}, \bibinfo {author} {\bibfnamefont {S.}~\bibnamefont {Shrestha}},
		\bibinfo {author} {\bibfnamefont {S.~C.}\ \bibnamefont {Malek}}, \bibinfo
		{author} {\bibfnamefont {M.}~\bibnamefont {Lu}}, \bibinfo {author}
		{\bibfnamefont {A.}~\bibnamefont {Stein}}, \bibinfo {author} {\bibfnamefont
			{C.}~\bibnamefont {Zheng}},\ and\ \bibinfo {author} {\bibfnamefont
			{N.}~\bibnamefont {Yu}},\ }\bibfield  {title} {\bibinfo {title} {Dielectric
			metasurfaces for complete and independent control of the optical amplitude
			and phase},\ }\href@noop {} {\bibfield  {journal} {\bibinfo  {journal} {Light
				Sci. Appl.}\ }\textbf {\bibinfo {volume} {8}},\ \bibinfo {pages} {92}
		(\bibinfo {year} {2019})}\BibitemShut {NoStop}%
	\bibitem [{\citenamefont {Zhang}\ \emph
		{et~al.}(2019{\natexlab{b}})\citenamefont {Zhang}, \citenamefont {Dong},
		\citenamefont {Intaravanne}, \citenamefont {Zang}, \citenamefont {Xu},
		\citenamefont {Song}, \citenamefont {Zheng}, \citenamefont {Wang},
		\citenamefont {Chu},\ and\ \citenamefont {Chen}}]{zhang2019multichannel}%
	\BibitemOpen
	\bibfield  {author} {\bibinfo {author} {\bibfnamefont {C.}~\bibnamefont
			{Zhang}}, \bibinfo {author} {\bibfnamefont {F.}~\bibnamefont {Dong}},
		\bibinfo {author} {\bibfnamefont {Y.}~\bibnamefont {Intaravanne}}, \bibinfo
		{author} {\bibfnamefont {X.}~\bibnamefont {Zang}}, \bibinfo {author}
		{\bibfnamefont {L.}~\bibnamefont {Xu}}, \bibinfo {author} {\bibfnamefont
			{Z.}~\bibnamefont {Song}}, \bibinfo {author} {\bibfnamefont {G.}~\bibnamefont
			{Zheng}}, \bibinfo {author} {\bibfnamefont {W.}~\bibnamefont {Wang}},
		\bibinfo {author} {\bibfnamefont {W.}~\bibnamefont {Chu}},\ and\ \bibinfo
		{author} {\bibfnamefont {X.}~\bibnamefont {Chen}},\ }\bibfield  {title}
	{\bibinfo {title} {Multichannel metasurfaces for anticounterfeiting},\
	}\href@noop {} {\bibfield  {journal} {\bibinfo  {journal} {Phys. Rev.
				Applied}\ }\textbf {\bibinfo {volume} {12}},\ \bibinfo {pages} {034028}
		(\bibinfo {year} {2019}{\natexlab{b}})}\BibitemShut {NoStop}%
	\bibitem [{\citenamefont {Hu}\ \emph {et~al.}(2019)\citenamefont {Hu},
		\citenamefont {Luo}, \citenamefont {Chen}, \citenamefont {Liu}, \citenamefont
		{Li}, \citenamefont {Wang}, \citenamefont {Liu},\ and\ \citenamefont
		{Duan}}]{hu20193d}%
	\BibitemOpen
	\bibfield  {author} {\bibinfo {author} {\bibfnamefont {Y.}~\bibnamefont
			{Hu}}, \bibinfo {author} {\bibfnamefont {X.}~\bibnamefont {Luo}}, \bibinfo
		{author} {\bibfnamefont {Y.}~\bibnamefont {Chen}}, \bibinfo {author}
		{\bibfnamefont {Q.}~\bibnamefont {Liu}}, \bibinfo {author} {\bibfnamefont
			{X.}~\bibnamefont {Li}}, \bibinfo {author} {\bibfnamefont {Y.}~\bibnamefont
			{Wang}}, \bibinfo {author} {\bibfnamefont {N.}~\bibnamefont {Liu}},\ and\
		\bibinfo {author} {\bibfnamefont {H.}~\bibnamefont {Duan}},\ }\bibfield
	{title} {\bibinfo {title} {3d-integrated metasurfaces for full-colour
			holography},\ }\href@noop {} {\bibfield  {journal} {\bibinfo  {journal}
			{Light Sci. Appl.}\ }\textbf {\bibinfo {volume} {8}},\ \bibinfo {pages} {86}
		(\bibinfo {year} {2019})}\BibitemShut {NoStop}%
	\bibitem [{\citenamefont {Dai}\ \emph {et~al.}(2020{\natexlab{b}})\citenamefont
		{Dai}, \citenamefont {Guan}, \citenamefont {Chang}, \citenamefont {Deng},
		\citenamefont {Tao}, \citenamefont {Li}, \citenamefont {Li}, \citenamefont
		{Yu}, \citenamefont {Zheng},\ and\ \citenamefont {Zhang}}]{dai2020single}%
	\BibitemOpen
	\bibfield  {author} {\bibinfo {author} {\bibfnamefont {Q.}~\bibnamefont
			{Dai}}, \bibinfo {author} {\bibfnamefont {Z.}~\bibnamefont {Guan}}, \bibinfo
		{author} {\bibfnamefont {S.}~\bibnamefont {Chang}}, \bibinfo {author}
		{\bibfnamefont {L.}~\bibnamefont {Deng}}, \bibinfo {author} {\bibfnamefont
			{J.}~\bibnamefont {Tao}}, \bibinfo {author} {\bibfnamefont {Z.}~\bibnamefont
			{Li}}, \bibinfo {author} {\bibfnamefont {Z.}~\bibnamefont {Li}}, \bibinfo
		{author} {\bibfnamefont {S.}~\bibnamefont {Yu}}, \bibinfo {author}
		{\bibfnamefont {G.}~\bibnamefont {Zheng}},\ and\ \bibinfo {author}
		{\bibfnamefont {S.}~\bibnamefont {Zhang}},\ }\bibfield  {title} {\bibinfo
		{title} {A single-celled tri-functional metasurface enabled with triple
			manipulations of light},\ }\href@noop {} {\bibfield  {journal} {\bibinfo
			{journal} {Adv. Funct. Mater.}\ }\textbf {\bibinfo {volume} {30}},\ \bibinfo
		{pages} {2003990} (\bibinfo {year} {2020}{\natexlab{b}})}\BibitemShut
	{NoStop}%
	\bibitem [{\citenamefont {Li}\ \emph {et~al.}(2020)\citenamefont {Li},
		\citenamefont {Chen}, \citenamefont {Guan}, \citenamefont {Tao},
		\citenamefont {Chang}, \citenamefont {Dai}, \citenamefont {Xiao},
		\citenamefont {Cui}, \citenamefont {Wang}, \citenamefont {Yu} \emph
		{et~al.}}]{li2020three}%
	\BibitemOpen
	\bibfield  {author} {\bibinfo {author} {\bibfnamefont {Z.}~\bibnamefont
			{Li}}, \bibinfo {author} {\bibfnamefont {C.}~\bibnamefont {Chen}}, \bibinfo
		{author} {\bibfnamefont {Z.}~\bibnamefont {Guan}}, \bibinfo {author}
		{\bibfnamefont {J.}~\bibnamefont {Tao}}, \bibinfo {author} {\bibfnamefont
			{S.}~\bibnamefont {Chang}}, \bibinfo {author} {\bibfnamefont
			{Q.}~\bibnamefont {Dai}}, \bibinfo {author} {\bibfnamefont {Y.}~\bibnamefont
			{Xiao}}, \bibinfo {author} {\bibfnamefont {Y.}~\bibnamefont {Cui}}, \bibinfo
		{author} {\bibfnamefont {Y.}~\bibnamefont {Wang}}, \bibinfo {author}
		{\bibfnamefont {S.}~\bibnamefont {Yu}}, \emph {et~al.},\ }\bibfield  {title}
	{\bibinfo {title} {Three-channel metasurfaces for simultaneous
			meta-holography and meta-nanoprinting: a single-cell design approach},\
	}\href@noop {} {\bibfield  {journal} {\bibinfo  {journal} {Laser Photonics
				Rev.}\ }\textbf {\bibinfo {volume} {14}},\ \bibinfo {pages} {2000032}
		(\bibinfo {year} {2020})}\BibitemShut {NoStop}%
	\bibitem [{\citenamefont {Zheng}\ \emph {et~al.}(2021)\citenamefont {Zheng},
		\citenamefont {Dai}, \citenamefont {Li}, \citenamefont {Ye}, \citenamefont
		{Xiong}, \citenamefont {Liu}, \citenamefont {Zheng},\ and\ \citenamefont
		{Zhang}}]{zheng2021metasurface}%
	\BibitemOpen
	\bibfield  {author} {\bibinfo {author} {\bibfnamefont {P.}~\bibnamefont
			{Zheng}}, \bibinfo {author} {\bibfnamefont {Q.}~\bibnamefont {Dai}}, \bibinfo
		{author} {\bibfnamefont {Z.}~\bibnamefont {Li}}, \bibinfo {author}
		{\bibfnamefont {Z.}~\bibnamefont {Ye}}, \bibinfo {author} {\bibfnamefont
			{J.}~\bibnamefont {Xiong}}, \bibinfo {author} {\bibfnamefont {H.-C.}\
			\bibnamefont {Liu}}, \bibinfo {author} {\bibfnamefont {G.}~\bibnamefont
			{Zheng}},\ and\ \bibinfo {author} {\bibfnamefont {S.}~\bibnamefont {Zhang}},\
	}\bibfield  {title} {\bibinfo {title} {Metasurface-based key for
			computational imaging encryption},\ }\href@noop {} {\bibfield  {journal}
		{\bibinfo  {journal} {Sci. Adv.}\ }\textbf {\bibinfo {volume} {7}},\ \bibinfo
		{pages} {eabg0363} (\bibinfo {year} {2021})}\BibitemShut {NoStop}%
	\bibitem [{\citenamefont {Yu}\ \emph {et~al.}(2024)\citenamefont {Yu},
		\citenamefont {Li}, \citenamefont {Zhao}, \citenamefont {Huang},
		\citenamefont {Lin}, \citenamefont {Yao}, \citenamefont {Li}, \citenamefont
		{Zhao}, \citenamefont {Wu}, \citenamefont {Li} \emph {et~al.}}]{yu2024high}%
	\BibitemOpen
	\bibfield  {author} {\bibinfo {author} {\bibfnamefont {Z.}~\bibnamefont
			{Yu}}, \bibinfo {author} {\bibfnamefont {H.}~\bibnamefont {Li}}, \bibinfo
		{author} {\bibfnamefont {W.}~\bibnamefont {Zhao}}, \bibinfo {author}
		{\bibfnamefont {P.-S.}\ \bibnamefont {Huang}}, \bibinfo {author}
		{\bibfnamefont {Y.-T.}\ \bibnamefont {Lin}}, \bibinfo {author} {\bibfnamefont
			{J.}~\bibnamefont {Yao}}, \bibinfo {author} {\bibfnamefont {W.}~\bibnamefont
			{Li}}, \bibinfo {author} {\bibfnamefont {Q.}~\bibnamefont {Zhao}}, \bibinfo
		{author} {\bibfnamefont {P.~C.}\ \bibnamefont {Wu}}, \bibinfo {author}
		{\bibfnamefont {B.}~\bibnamefont {Li}}, \emph {et~al.},\ }\bibfield  {title}
	{\bibinfo {title} {High-security learning-based optical encryption assisted
			by disordered metasurface},\ }\href@noop {} {\bibfield  {journal} {\bibinfo
			{journal} {Nat. Commun.}\ }\textbf {\bibinfo {volume} {15}},\ \bibinfo
		{pages} {2607} (\bibinfo {year} {2024})}\BibitemShut {NoStop}%
	\bibitem [{\citenamefont {Cao}\ \emph {et~al.}(2022)\citenamefont {Cao},
		\citenamefont {Tang}, \citenamefont {Li}, \citenamefont {Lee},\ and\
		\citenamefont {Dong}}]{cao2022four}%
	\BibitemOpen
	\bibfield  {author} {\bibinfo {author} {\bibfnamefont {Y.}~\bibnamefont
			{Cao}}, \bibinfo {author} {\bibfnamefont {L.}~\bibnamefont {Tang}}, \bibinfo
		{author} {\bibfnamefont {J.}~\bibnamefont {Li}}, \bibinfo {author}
		{\bibfnamefont {C.}~\bibnamefont {Lee}},\ and\ \bibinfo {author}
		{\bibfnamefont {Z.-G.}\ \bibnamefont {Dong}},\ }\bibfield  {title} {\bibinfo
		{title} {Four-channel display and encryption by near-field reflection on
			nanoprinting metasurface},\ }\href@noop {} {\bibfield  {journal} {\bibinfo
			{journal} {Nanophotonics}\ }\textbf {\bibinfo {volume} {11}},\ \bibinfo
		{pages} {3365} (\bibinfo {year} {2022})}\BibitemShut {NoStop}%
	\bibitem [{\citenamefont {Xing}\ \emph {et~al.}(2025)\citenamefont {Xing},
		\citenamefont {Bu}, \citenamefont {Zhang}, \citenamefont {Choi},
		\citenamefont {Li}, \citenamefont {Yue}, \citenamefont {Cheng}, \citenamefont
		{Li}, \citenamefont {Chen},\ and\ \citenamefont {Gao}}]{xing2025metasurface}%
	\BibitemOpen
	\bibfield  {author} {\bibinfo {author} {\bibfnamefont {W.}~\bibnamefont
			{Xing}}, \bibinfo {author} {\bibfnamefont {C.}~\bibnamefont {Bu}}, \bibinfo
		{author} {\bibfnamefont {X.}~\bibnamefont {Zhang}}, \bibinfo {author}
		{\bibfnamefont {D.-Y.}\ \bibnamefont {Choi}}, \bibinfo {author}
		{\bibfnamefont {Y.}~\bibnamefont {Li}}, \bibinfo {author} {\bibfnamefont
			{W.}~\bibnamefont {Yue}}, \bibinfo {author} {\bibfnamefont {J.}~\bibnamefont
			{Cheng}}, \bibinfo {author} {\bibfnamefont {Z.}~\bibnamefont {Li}}, \bibinfo
		{author} {\bibfnamefont {S.}~\bibnamefont {Chen}},\ and\ \bibinfo {author}
		{\bibfnamefont {S.}~\bibnamefont {Gao}},\ }\bibfield  {title} {\bibinfo
		{title} {Metasurface-enabled optical encryption and steganography with
			enhanced information security},\ }\href@noop {} {\bibfield  {journal}
		{\bibinfo  {journal} {Nanophotonics}\ }\textbf {\bibinfo {volume} {14}},\
		\bibinfo {pages} {1391} (\bibinfo {year} {2025})}\BibitemShut {NoStop}%
	\bibitem [{\citenamefont {Zhu}\ \emph {et~al.}(2025)\citenamefont {Zhu},
		\citenamefont {Xie}, \citenamefont {Dong}, \citenamefont {Shang},
		\citenamefont {Guan}, \citenamefont {Burokur},\ and\ \citenamefont
		{Ding}}]{zhu2025metasurface}%
	\BibitemOpen
	\bibfield  {author} {\bibinfo {author} {\bibfnamefont {L.}~\bibnamefont
			{Zhu}}, \bibinfo {author} {\bibfnamefont {W.}~\bibnamefont {Xie}}, \bibinfo
		{author} {\bibfnamefont {L.}~\bibnamefont {Dong}}, \bibinfo {author}
		{\bibfnamefont {G.}~\bibnamefont {Shang}}, \bibinfo {author} {\bibfnamefont
			{C.}~\bibnamefont {Guan}}, \bibinfo {author} {\bibfnamefont {S.~N.}\
			\bibnamefont {Burokur}},\ and\ \bibinfo {author} {\bibfnamefont
			{X.}~\bibnamefont {Ding}},\ }\bibfield  {title} {\bibinfo {title}
		{Metasurface-assisted chaotic cryptography platform for enhanced secure
			communication},\ }\href@noop {} {\bibfield  {journal} {\bibinfo  {journal}
			{Laser Photonics Rev.}\ ,\ \bibinfo {pages} {e00955}} (\bibinfo {year}
		{2025})}\BibitemShut {NoStop}%
	\bibitem [{\citenamefont {Levy}(1994)}]{levy1994chaos}%
	\BibitemOpen
	\bibfield  {author} {\bibinfo {author} {\bibfnamefont {D.}~\bibnamefont
			{Levy}},\ }\bibfield  {title} {\bibinfo {title} {Chaos theory and strategy:
			Theory, application, and managerial implications},\ }\href@noop {} {\bibfield
		{journal} {\bibinfo  {journal} {Strategic Manage. J.}\ }\textbf {\bibinfo
			{volume} {15}},\ \bibinfo {pages} {167} (\bibinfo {year} {1994})}\BibitemShut
	{NoStop}%
	\bibitem [{\citenamefont {Chen}\ \emph {et~al.}(2015)\citenamefont {Chen},
		\citenamefont {Zhu}, \citenamefont {Fu}, \citenamefont {Yu},\ and\
		\citenamefont {Zhang}}]{chen2015fast}%
	\BibitemOpen
	\bibfield  {author} {\bibinfo {author} {\bibfnamefont {J.-x.}\ \bibnamefont
			{Chen}}, \bibinfo {author} {\bibfnamefont {Z.-l.}\ \bibnamefont {Zhu}},
		\bibinfo {author} {\bibfnamefont {C.}~\bibnamefont {Fu}}, \bibinfo {author}
		{\bibfnamefont {H.}~\bibnamefont {Yu}},\ and\ \bibinfo {author}
		{\bibfnamefont {L.-b.}\ \bibnamefont {Zhang}},\ }\bibfield  {title} {\bibinfo
		{title} {A fast chaos-based image encryption scheme with a dynamic state
			variables selection mechanism},\ }\href@noop {} {\bibfield  {journal}
		{\bibinfo  {journal} {Commun. Nonlinear Sci.}\ }\textbf {\bibinfo {volume}
			{20}},\ \bibinfo {pages} {846} (\bibinfo {year} {2015})}\BibitemShut
	{NoStop}%
	\bibitem [{\citenamefont {Yoon}\ and\ \citenamefont
		{Kim}(2010)}]{yoon2010image}%
	\BibitemOpen
	\bibfield  {author} {\bibinfo {author} {\bibfnamefont {J.~W.}\ \bibnamefont
			{Yoon}}\ and\ \bibinfo {author} {\bibfnamefont {H.}~\bibnamefont {Kim}},\
	}\bibfield  {title} {\bibinfo {title} {An image encryption scheme with a
			pseudorandom permutation based on chaotic maps},\ }\href@noop {} {\bibfield
		{journal} {\bibinfo  {journal} {Commun. Nonlinear Sci.}\ }\textbf {\bibinfo
			{volume} {15}},\ \bibinfo {pages} {3998} (\bibinfo {year}
		{2010})}\BibitemShut {NoStop}%
	\bibitem [{\citenamefont {Zhu}\ \emph {et~al.}(2011)\citenamefont {Zhu},
		\citenamefont {Zhang}, \citenamefont {Wong},\ and\ \citenamefont
		{Yu}}]{zhu2011chaos}%
	\BibitemOpen
	\bibfield  {author} {\bibinfo {author} {\bibfnamefont {Z.-l.}\ \bibnamefont
			{Zhu}}, \bibinfo {author} {\bibfnamefont {W.}~\bibnamefont {Zhang}}, \bibinfo
		{author} {\bibfnamefont {K.-w.}\ \bibnamefont {Wong}},\ and\ \bibinfo
		{author} {\bibfnamefont {H.}~\bibnamefont {Yu}},\ }\bibfield  {title}
	{\bibinfo {title} {A chaos-based symmetric image encryption scheme using a
			bit-level permutation},\ }\href@noop {} {\bibfield  {journal} {\bibinfo
			{journal} {Inform. Sciences}\ }\textbf {\bibinfo {volume} {181}},\ \bibinfo
		{pages} {1171} (\bibinfo {year} {2011})}\BibitemShut {NoStop}%
	\bibitem [{\citenamefont {Zhou}\ \emph {et~al.}(2024)\citenamefont {Zhou},
		\citenamefont {Xia}, \citenamefont {Lin}, \citenamefont {Yang}, \citenamefont
		{Zhang},\ and\ \citenamefont {Zhou}}]{zhou2024two}%
	\BibitemOpen
	\bibfield  {author} {\bibinfo {author} {\bibfnamefont {L.}~\bibnamefont
			{Zhou}}, \bibinfo {author} {\bibfnamefont {H.}~\bibnamefont {Xia}}, \bibinfo
		{author} {\bibfnamefont {Q.}~\bibnamefont {Lin}}, \bibinfo {author}
		{\bibfnamefont {X.}~\bibnamefont {Yang}}, \bibinfo {author} {\bibfnamefont
			{X.}~\bibnamefont {Zhang}},\ and\ \bibinfo {author} {\bibfnamefont
			{M.}~\bibnamefont {Zhou}},\ }\bibfield  {title} {\bibinfo {title}
		{Two-dimensional hyperchaos-based encryption and compression algorithm for
			agricultural uav-captured planar images},\ }\href@noop {} {\bibfield
		{journal} {\bibinfo  {journal} {Sci. Rep.}\ }\textbf {\bibinfo {volume}
			{14}},\ \bibinfo {pages} {22423} (\bibinfo {year} {2024})}\BibitemShut
	{NoStop}%
	\bibitem [{\citenamefont {Li}\ \emph {et~al.}(2025{\natexlab{b}})\citenamefont
		{Li}, \citenamefont {Wan}, \citenamefont {Xiao},\ and\ \citenamefont
		{Liu}}]{li2025tunable}%
	\BibitemOpen
	\bibfield  {author} {\bibinfo {author} {\bibfnamefont {Y.}~\bibnamefont
			{Li}}, \bibinfo {author} {\bibfnamefont {Z.}~\bibnamefont {Wan}}, \bibinfo
		{author} {\bibfnamefont {S.}~\bibnamefont {Xiao}},\ and\ \bibinfo {author}
		{\bibfnamefont {T.}~\bibnamefont {Liu}},\ }\bibfield  {title} {\bibinfo
		{title} {Tunable phase-change metasurfaces for integrated display and
			encryption: dual near-field nanoprinting and single far-field holography},\
	}\href@noop {} {\bibfield  {journal} {\bibinfo  {journal} {J. Opt.}\ }\textbf
		{\bibinfo {volume} {27}},\ \bibinfo {pages} {065102} (\bibinfo {year}
		{2025}{\natexlab{b}})}\BibitemShut {NoStop}%
	\bibitem [{\citenamefont {Liu}\ \emph {et~al.}(2022)\citenamefont {Liu},
		\citenamefont {Han}, \citenamefont {Duan},\ and\ \citenamefont
		{Xiao}}]{liu2022phase}%
	\BibitemOpen
	\bibfield  {author} {\bibinfo {author} {\bibfnamefont {T.}~\bibnamefont
			{Liu}}, \bibinfo {author} {\bibfnamefont {Z.}~\bibnamefont {Han}}, \bibinfo
		{author} {\bibfnamefont {J.}~\bibnamefont {Duan}},\ and\ \bibinfo {author}
		{\bibfnamefont {S.}~\bibnamefont {Xiao}},\ }\bibfield  {title} {\bibinfo
		{title} {Phase-change metasurfaces for dynamic image display and information
			encryption},\ }\href@noop {} {\bibfield  {journal} {\bibinfo  {journal}
			{Phys. Rev. Applied}\ }\textbf {\bibinfo {volume} {18}},\ \bibinfo {pages}
		{044078} (\bibinfo {year} {2022})}\BibitemShut {NoStop}%
\end{thebibliography}

%

\end{document}